\documentclass [12pt,a4paper]{article}
\def \lsim{\mathrel{\mathpalette\@versim<}}
\def \gsim{\mathrel{\mathpalette\@versim>}}

\def\sba2{\sin ^2 (\beta - \alpha)}
\def\cba2{\cos ^2 (\beta - \alpha)}

\def\r{\rightarrow}

\def\sqs{\sqrt{s}}

\def\W{\mathrm W^{\pm}}
\def\Z{\mathrm Z}

\def\H{\mathrm H}

\def\ee{\mathrm e^+\mathrm e^-}
\def\mm{\mu^{+}\mu^{-}}
\def\nn{\nu \bar{\nu}}
\def\qq{\mathrm q \bar{\mathrm q}}
\def\pb{ ~\mathrm{pb} ^{-1}}

\def\Gc{\mathrm{~GeV}/c}
\def\G{\mathrm{~GeV}}

\def\tt{\tau^{+}\tau^{-}}

\def\ll{\ell^{+}\ell^{-}}

\def\bb{\mathrm b \bar{\mathrm b}}
\def\mtau{m_{\tau}}

\def\mH{m_{\H}}

\def\mZ{m_{\Z}}

\def\pt{p_{\mathrm{t}}}
\def\G{ \rm{GeV} }

\newcommand{\mvis}{\ensuremath{m_{\mathrm{vis}}}}
\newcommand{\ahz}{\ensuremath{{\tilde{A}}}}
\newcommand{\pmiss}{\ensuremath{{\not p}}}
\newcommand{\pms}{\ensuremath{{\theta_{\pmiss}}}}

\newcommand{\emj}{\ensuremath{{E_\tau}}}

\newcommand{\ZZ}{\ensuremath{{\mathrm Z}{\mathrm Z}}}
\newcommand{\qqg}{\ensuremath{{\mathrm q}\bar{\mathrm q}(\gamma)}}
\newcommand{\WW}{\ensuremath{{\mathrm W}^{+}{\mathrm W}^{-}}}
\newcommand{\Gcs}{\ensuremath{{\mathrm{~GeV}/c^2}}}
\newcommand{\HZS}{\ensuremath{\H\Z\r\bb\qq}}

\usepackage{epsfig}
\begin{document}
\pagestyle{empty}

\title{Search for the Standard Model Higgs Boson \\ at the LEP2 Collider
near $\protect\sqrt{s}= 183 ~\G$
\vspace{1cm}}
\author{The {\sc ALEPH} Collaboration$^*)$}
\date{\mbox{ }}
\maketitle

\begin{picture}(160,1)
\put(19,110){\rm EUROPEAN LABORATORY FOR PARTICLE PHYSICS (CERN)}
%\put(-5,94){\parbox[t]{45mm}{FINAL DRAFT }}
\put(125,94){\parbox[t]{45mm}{CERN-EP/98-144}}
\put(125,88){\parbox[t]{45mm}{September 16,1998}}
\end{picture}

\vspace{1cm}
\begin{abstract}
\vspace{.5cm}
During 1997 the ALEPH experiment at LEP
%the Large Electron Positron collider
gathered $57 \pb$ of data at centre-of-mass energies near $183 ~\G$.
These data are used to look for possible signals from the production
of the Standard Model Higgs boson in the reaction $\ee\r\H\Z$.
%The search addresses the following channels: $\H\ee$ and
%$\H\mm$, $\H\nn$, $\H\qq$, $\H\tt$ and $\tt\qq$.
%In this letter we describe updates of the previously published event
%selections as well as new event selections -- based on neural networks
%-- which are also used to derive the result.
No evidence of a signal is found in the data; seven events are
selected, in agreement with the expectation of 7.2 events from
background processes.  This observation results in an improved lower
limit on the mass of the Higgs boson: $\mH > 87.9 \Gcs$ at 95\%
confidence level.
\end{abstract}

\vspace{1.5cm}
\begin{center}
The ALEPH Collaboration wish to dedicate this paper to \\
the memory of Colin Raine who died suddenly on September 16, 1998.
%This paper is dedicated to the memory of Colin Raine,\\
%who died on September 16th, 1998.
\end{center}

\vfill
%\centerline{{\sl Submitted to Physics Letters B}}
\centerline{{\sl Accepted for publication in Physics Letters B}}
\vskip .5cm
\noindent
--------------------------------------------\hfil\break
{\small $^*)$ See next pages for the list of authors}
%
% Author list
%
\eject
\setlength{\topmargin}{-1cm}
\setlength{\oddsidemargin}{-0.5cm}
%------------------------------------------------------------------------
% authob12pt.tex
% authors' list for papers at LEP 1.5 and 2 energies
%-----------------------------------------------------------------------
\pagestyle{empty}
\newpage
\small
%
% remember the old settings
%
\newlength{\saveparskip}
\newlength{\savetextheight}
\newlength{\savetopmargin}
\newlength{\savetextwidth}
\newlength{\saveoddsidemargin}
\newlength{\savetopsep}
\setlength{\saveparskip}{\parskip}
\setlength{\savetextheight}{\textheight}
\setlength{\savetopmargin}{\topmargin}
\setlength{\savetextwidth}{\textwidth}
\setlength{\saveoddsidemargin}{\oddsidemargin}
\setlength{\savetopsep}{\topsep}
%
% text dimensions for the author list
%
\setlength{\parskip}{0.0cm}
\setlength{\textheight}{25.0cm}
\setlength{\textwidth}{16 cm}
\setlength{\topsep}{1mm}
\pretolerance=10000
%%%%%%%%%\begin{document}
%\centerline{EUROPEAN ORGANIZATION FOR NUCLEAR RESEARCH}
%\centerline{EUROPEAN LABORATORY FOR PARTICLE PHYSICS (CERN)}
%\vspace{1cm}
%\begin{flushright}CERN-PPE/96-   \\
%5 October 1998 - last update
%\end{flushright}
\centerline{\large\bf The ALEPH Collaboration}
\footnotesize
\vspace{0.5cm}
{\raggedbottom
\begin{sloppypar}
\samepage\noindent
R.~Barate,
D.~Buskulic,
D.~Decamp,
P.~Ghez,
C.~Goy,
S.~Jezequel,
J.-P.~Lees,
A.~Lucotte,
F.~Martin,
E.~Merle,
\mbox{M.-N.~Minard},
\mbox{J.-Y.~Nief},
P.~Perrodo,
B.~Pietrzyk
\nopagebreak
\begin{center}
\parbox{15.5cm}{\sl\samepage
Laboratoire de Physique des Particules (LAPP), IN$^{2}$P$^{3}$-CNRS,
F-74019 Annecy-le-Vieux Cedex, France}
\end{center}\end{sloppypar}
\vspace{2mm}
\begin{sloppypar}
\noindent
R.~Alemany,
M.P.~Casado,
M.~Chmeissani,
J.M.~Crespo,
M.~Delfino,
E.~Fernandez,
M.~Fernandez-Bosman,
Ll.~Garrido,$^{15}$
E.~Graug\`{e}s,
A.~Juste,
M.~Martinez,
G.~Merino,
R.~Miquel,
Ll.M.~Mir,
P.~Morawitz,
A.~Pacheco,
I.C.~Park,
A.~Pascual,
I.~Riu,
F.~Sanchez
\nopagebreak
\begin{center}
\parbox{15.5cm}{\sl\samepage
Institut de F\'{i}sica d'Altes Energies, Universitat Aut\`{o}noma
de Barcelona, 08193 Bellaterra (Barcelona), E-Spain$^{7}$}
\end{center}\end{sloppypar}
\vspace{2mm}
\begin{sloppypar}
\noindent
A.~Colaleo,
D.~Creanza,
M.~de~Palma,
G.~Gelao,
G.~Iaselli,
G.~Maggi,
M.~Maggi,
S.~Nuzzo,
A.~Ranieri,
G.~Raso,
F.~Ruggieri,
G.~Selvaggi,
L.~Silvestris,
P.~Tempesta,
A.~Tricomi,$^{3}$
G.~Zito
\nopagebreak
\begin{center}
\parbox{15.5cm}{\sl\samepage
Dipartimento di Fisica, INFN Sezione di Bari, I-70126 Bari, Italy}
\end{center}\end{sloppypar}
\vspace{2mm}
\begin{sloppypar}
\noindent
X.~Huang,
J.~Lin,
Q. Ouyang,
T.~Wang,
Y.~Xie,
R.~Xu,
S.~Xue,
J.~Zhang,
L.~Zhang,
W.~Zhao
\nopagebreak
\begin{center}
\parbox{15.5cm}{\sl\samepage
Institute of High-Energy Physics, Academia Sinica, Beijing, The People's
Republic of China$^{8}$}
\end{center}\end{sloppypar}
\vspace{2mm}
\begin{sloppypar}
\noindent
D.~Abbaneo,
U.~Becker,$^{22}$
G.~Boix,$^{24}$
M.~Cattaneo,
F.~Cerutti,
V.~Ciulli,
G.~Dissertori,
H.~Drevermann,
R.W.~Forty,
M.~Frank,
F.~Gianotti,
R.~Hagelberg,
A.W.~Halley,
J.B.~Hansen,
J.~Harvey,
P.~Janot,
B.~Jost,
I.~Lehraus,
O.~Leroy,
P.~Maley,
P.~Mato,
A.~Minten,
L.~Moneta,$^{20}$
A.~Moutoussi,
F.~Ranjard,
L.~Rolandi,
D.~Rousseau,
D.~Schlatter,
M.~Schmitt,$^{1}$
O.~Schneider,
W.~Tejessy,
F.~Teubert,
I.R.~Tomalin,
E.~Tournefier,
M.~Vreeswijk,
H.~Wachsmuth
\nopagebreak
\begin{center}
\parbox{15.5cm}{\sl\samepage
European Laboratory for Particle Physics (CERN), CH-1211 Geneva 23,
Switzerland}
\end{center}\end{sloppypar}
\vspace{2mm}
\begin{sloppypar}
\noindent
Z.~Ajaltouni,
F.~Badaud
G.~Chazelle,
O.~Deschamps,
S.~Dessagne,
A.~Falvard,
C.~Ferdi,
P.~Gay,
C.~Guicheney,
P.~Henrard,
J.~Jousset,
B.~Michel,
S.~Monteil,
\mbox{J-C.~Montret},
D.~Pallin,
P.~Perret,
F.~Podlyski
\nopagebreak
\begin{center}
\parbox{15.5cm}{\sl\samepage
Laboratoire de Physique Corpusculaire, Universit\'e Blaise Pascal,
IN$^{2}$P$^{3}$-CNRS, Clermont-Ferrand, F-63177 Aubi\`{e}re, France}
\end{center}\end{sloppypar}
\vspace{2mm}
\begin{sloppypar}
\noindent
J.D.~Hansen,
J.R.~Hansen,
P.H.~Hansen,
B.S.~Nilsson,
B.~Rensch,
A.~W\"a\"an\"anen
\begin{center}
\parbox{15.5cm}{\sl\samepage
Niels Bohr Institute, 2100 Copenhagen, DK-Denmark$^{9}$}
\end{center}\end{sloppypar}
\vspace{2mm}
\begin{sloppypar}
\noindent
G.~Daskalakis,
A.~Kyriakis,
C.~Markou,
E.~Simopoulou,
A.~Vayaki
\nopagebreak
\begin{center}
\parbox{15.5cm}{\sl\samepage
Nuclear Research Center Demokritos (NRCD), GR-15310 Attiki, Greece}
\end{center}\end{sloppypar}
\vspace{2mm}
\begin{sloppypar}
\noindent
A.~Blondel,
\mbox{J.-C.~Brient},
F.~Machefert,
A.~Roug\'{e},
M.~Rumpf,
R.~Tanaka,
A.~Valassi,$^{6}$
H.~Videau
\nopagebreak
\begin{center}
\parbox{15.5cm}{\sl\samepage
Laboratoire de Physique Nucl\'eaire et des Hautes Energies, Ecole
Polytechnique, IN$^{2}$P$^{3}$-CNRS, \mbox{F-91128} Palaiseau Cedex, France}
\end{center}\end{sloppypar}
\vspace{2mm}
\begin{sloppypar}
\noindent
E.~Focardi,
G.~Parrini,
K.~Zachariadou
\nopagebreak
\begin{center}
\parbox{15.5cm}{\sl\samepage
Dipartimento di Fisica, Universit\`a di Firenze, INFN Sezione di Firenze,
I-50125 Firenze, Italy}
\end{center}\end{sloppypar}
\vspace{2mm}
\begin{sloppypar}
\noindent
R.~Cavanaugh,
M.~Corden,
C.~Georgiopoulos,
T.~Huehn,
D.E.~Jaffe
\nopagebreak
\begin{center}
\parbox{15.5cm}{\sl\samepage
Supercomputer Computations Research Institute,
Florida State University,
Tallahassee, FL 32306-4052, USA $^{13,14}$}
\end{center}\end{sloppypar}
\vspace{2mm}
\begin{sloppypar}
\noindent
A.~Antonelli,
G.~Bencivenni,
G.~Bologna,$^{4}$
F.~Bossi,
P.~Campana,
G.~Capon,
V.~Chiarella,
P.~Laurelli,
G.~Mannocchi,$^{5}$
F.~Murtas,
G.P.~Murtas,
L.~Passalacqua,
M.~Pepe-Altarelli$^{12}$
\nopagebreak
\begin{center}
\parbox{15.5cm}{\sl\samepage
Laboratori Nazionali dell'INFN (LNF-INFN), I-00044 Frascati, Italy}
\end{center}\end{sloppypar}
\vspace{2mm}
%\pagebreak
\begin{sloppypar}
\noindent
M.~Chalmers,
L.~Curtis,
J.G.~Lynch,
P.~Negus,
V.~O'Shea,
C.~Raine,
J.M.~Scarr,
P.~Teixeira-Dias,
A.S.~Thompson,
E.~Thomson,
J.J.~Ward
\nopagebreak
\begin{center}
\parbox{15.5cm}{\sl\samepage
Department of Physics and Astronomy, University of Glasgow, Glasgow G12
8QQ,United Kingdom$^{10}$}
\end{center}\end{sloppypar}
%\vspace{2mm}
\pagebreak
\begin{sloppypar}
\noindent
O.~Buchm\"uller,
S.~Dhamotharan,
C.~Geweniger,
P.~Hanke,
G.~Hansper,
V.~Hepp,
E.E.~Kluge,
A.~Putzer,
J.~Sommer,
K.~Tittel,
S.~Werner,
M.~Wunsch
\nopagebreak
\begin{center}
\parbox{15.5cm}{\sl\samepage
Institut f\"ur Hochenergiephysik, Universit\"at Heidelberg, D-69120
Heidelberg, Germany$^{16}$}
\end{center}\end{sloppypar}
\vspace{2mm}
\begin{sloppypar}
\noindent
R.~Beuselinck,
D.M.~Binnie,
W.~Cameron,
P.J.~Dornan,$^{12}$
M.~Girone,
S.~Goodsir,
N.~Marinelli,
E.B.~Martin,
J.~Nash,
J.K.~Sedgbeer,
P.~Spagnolo,
M.D.~Williams
\nopagebreak
\begin{center}
\parbox{15.5cm}{\sl\samepage
Department of Physics, Imperial College, London SW7 2BZ,
United Kingdom$^{10}$}
\end{center}\end{sloppypar}
\vspace{2mm}
\begin{sloppypar}
\noindent
V.M.~Ghete,
P.~Girtler,
E.~Kneringer,
D.~Kuhn,
G.~Rudolph
\nopagebreak
\begin{center}
\parbox{15.5cm}{\sl\samepage
Institut f\"ur Experimentalphysik, Universit\"at Innsbruck, A-6020
Innsbruck, Austria$^{18}$}
\end{center}\end{sloppypar}
\vspace{2mm}
\begin{sloppypar}
\noindent
A.P.~Betteridge,
C.K.~Bowdery,
P.G.~Buck,
P.~Colrain,
G.~Crawford,
G.~Ellis,
A.J.~Finch,
F.~Foster,
G.~Hughes,
R.W.L.~Jones,
A.N.~Robertson,
M.I.~Williams
\nopagebreak
\begin{center}
\parbox{15.5cm}{\sl\samepage
Department of Physics, University of Lancaster, Lancaster LA1 4YB,
United Kingdom$^{10}$}
\end{center}\end{sloppypar}
\vspace{2mm}
\begin{sloppypar}
\noindent
P.~van~Gemmeren,
I.~Giehl,
C.~Hoffmann,
K.~Jakobs,
K.~Kleinknecht,
M.~Kr\"ocker,
H.-A.~N\"urnberger,
G.~Quast,
B.~Renk,
E.~Rohne,
H.-G.~Sander,
S.~Schmeling,
C.~Zeitnitz,
T.~Ziegler
\nopagebreak
\begin{center}
\parbox{15.5cm}{\sl\samepage
Institut f\"ur Physik, Universit\"at Mainz, D-55099 Mainz, Germany$^{16}$}
\end{center}\end{sloppypar}
\vspace{2mm}
\begin{sloppypar}
\noindent
J.J.~Aubert,
C.~Benchouk,
A.~Bonissent,
J.~Carr,$^{12}$
P.~Coyle,
A.~Ealet,
D.~Fouchez,
F.~Motsch,
P.~Payre,
M.~Talby,
M.~Thulasidas,
A.~Tilquin
\nopagebreak
\begin{center}
\parbox{15.5cm}{\sl\samepage
Centre de Physique des Particules, Facult\'e des Sciences de Luminy,
IN$^{2}$P$^{3}$-CNRS, F-13288 Marseille, France}
\end{center}\end{sloppypar}
\vspace{2mm}
\begin{sloppypar}
\noindent
M.~Aleppo,
M.~Antonelli,
F.~Ragusa
\nopagebreak
\begin{center}
\parbox{15.5cm}{\sl\samepage
Dipartimento di Fisica, Universit\`a di Milano e INFN Sezione di
Milano, I-20133 Milano, Italy.}
\end{center}\end{sloppypar}
\vspace{2mm}
\begin{sloppypar}
\noindent
R.~Berlich,
V.~B\"uscher,
H.~Dietl,
G.~Ganis,
K.~H\"uttmann,
G.~L\"utjens,
C.~Mannert,
W.~M\"anner,
\mbox{H.-G.~Moser},
S.~Schael,
R.~Settles,
H.~Seywerd,
H.~Stenzel,
W.~Wiedenmann,
G.~Wolf
\nopagebreak
\begin{center}
\parbox{15.5cm}{\sl\samepage
Max-Planck-Institut f\"ur Physik, Werner-Heisenberg-Institut,
D-80805 M\"unchen, Germany\footnotemark[16]}
\end{center}\end{sloppypar}
\vspace{2mm}
\begin{sloppypar}
\noindent
J.~Boucrot,
O.~Callot,
S.~Chen,
M.~Davier,
L.~Duflot,
\mbox{J.-F.~Grivaz},
Ph.~Heusse,
A.~H\"ocker,
A.~Jacholkowska,
M.~Kado,
D.W.~Kim,$^{2}$
F.~Le~Diberder,
J.~Lefran\c{c}ois,
L.~Serin,
\mbox{J.-J.~Veillet},
I.~Videau,$^{12}$
\hbox{J.-B}.~de~Vivie~de~R\'egie,
D.~Zerwas
\nopagebreak
\begin{center}
\parbox{15.5cm}{\sl\samepage
Laboratoire de l'Acc\'el\'erateur Lin\'eaire, Universit\'e de Paris-Sud,
IN$^{2}$P$^{3}$-CNRS, F-91898 Orsay Cedex, France}
\end{center}\end{sloppypar}
\vspace{2mm}
\begin{sloppypar}
\noindent
\samepage
P.~Azzurri,
G.~Bagliesi,$^{12}$
S.~Bettarini,
T.~Boccali,
C.~Bozzi,
G.~Calderini,
R.~Dell'Orso,
R.~Fantechi,
I.~Ferrante,
A.~Giassi,
A.~Gregorio,
F.~Ligabue,
A.~Lusiani,
P.S.~Marrocchesi,
A.~Messineo,
F.~Palla,
G.~Rizzo,
G.~Sanguinetti,
A.~Sciab\`a,
G.~Sguazzoni,
R.~Tenchini,
C.~Vannini,
A.~Venturi,
P.G.~Verdini
\samepage
\begin{center}
\parbox{15.5cm}{\sl\samepage
Dipartimento di Fisica dell'Universit\`a, INFN Sezione di Pisa,
e Scuola Normale Superiore, I-56010 Pisa, Italy}
\end{center}\end{sloppypar}
\vspace{2mm}
\begin{sloppypar}
\noindent
G.A.~Blair,
J.T.~Chambers,
J.~Coles,
G.~Cowan,
M.G.~Green,
T.~Medcalf,
J.A.~Strong,
J.H.~von~Wimmersperg-Toeller
\nopagebreak
\begin{center}
\parbox{15.5cm}{\sl\samepage
Department of Physics, Royal Holloway \& Bedford New College,
University of London, Surrey TW20 OEX, United Kingdom$^{10}$}
\end{center}\end{sloppypar}
\vspace{2mm}
\begin{sloppypar}
\noindent
D.R.~Botterill,
R.W.~Clifft,
T.R.~Edgecock,
P.R.~Norton,
J.C.~Thompson,
A.E.~Wright
\nopagebreak
\begin{center}
\parbox{15.5cm}{\sl\samepage
Particle Physics Dept., Rutherford Appleton Laboratory,
Chilton, Didcot, Oxon OX11 OQX, United Kingdom$^{10}$}
\end{center}\end{sloppypar}
\vspace{2mm}
%\pagebreak
\begin{sloppypar}
\noindent
\mbox{B.~Bloch-Devaux},
P.~Colas,
B.~Fabbro,
G.~Fa\"\i f,
E.~Lan\c{c}on,$^{12}$
\mbox{M.-C.~Lemaire},
E.~Locci,
P.~Perez,
H.~Przysiezniak,
J.~Rander,
\mbox{J.-F.~Renardy},
A.~Rosowsky,
A.~Trabelsi,$^{23}$
B.~Tuchming,
B.~Vallage
\nopagebreak
\begin{center}
\parbox{15.5cm}{\sl\samepage
CEA, DAPNIA/Service de Physique des Particules,
CE-Saclay, F-91191 Gif-sur-Yvette Cedex, France$^{17}$}
\end{center}\end{sloppypar}
%\nopagebreak
\vspace{2mm}
\begin{sloppypar}
\noindent
S.N.~Black,
J.H.~Dann,
H.Y.~Kim,
N.~Konstantinidis,
A.M.~Litke,
M.A. McNeil,
G.~Taylor
\nopagebreak
\begin{center}
\parbox{15.5cm}{\sl\samepage
Institute for Particle Physics, University of California at
Santa Cruz, Santa Cruz, CA 95064, USA$^{19}$}
\end{center}\end{sloppypar}
\pagebreak
\vspace{2mm}
\begin{sloppypar}
\noindent
C.N.~Booth,
S.~Cartwright,
F.~Combley,
M.S.~Kelly,
M.~Lehto,
L.F.~Thompson
\nopagebreak
\begin{center}
\parbox{15.5cm}{\sl\samepage
Department of Physics, University of Sheffield, Sheffield S3 7RH,
United Kingdom$^{10}$}
\end{center}\end{sloppypar}
\vspace{2mm}
\begin{sloppypar}
\noindent
K.~Affholderbach,
A.~B\"ohrer,
S.~Brandt,
J.~Foss,
C.~Grupen,
G.~Prange,
L.~Smolik,
F.~Stephan
\nopagebreak
\begin{center}
\parbox{15.5cm}{\sl\samepage
Fachbereich Physik, Universit\"at Siegen, D-57068 Siegen, Germany$^{16}$}
\end{center}\end{sloppypar}
\vspace{2mm}
\begin{sloppypar}
\noindent
G.~Giannini,
B.~Gobbo
\nopagebreak
\begin{center}
\parbox{15.5cm}{\sl\samepage
Dipartimento di Fisica, Universit\`a di Trieste e INFN Sezione di Trieste,
I-34127 Trieste, Italy}
\end{center}\end{sloppypar}
\vspace{2mm}
\begin{sloppypar}
\noindent
J.~Putz,
J.~Rothberg,
S.~Wasserbaech,
R.W.~Williams
\nopagebreak
\begin{center}
\parbox{15.5cm}{\sl\samepage
Experimental Elementary Particle Physics, University of Washington, WA 98195
Seattle, U.S.A.}
\end{center}\end{sloppypar}
\vspace{2mm}
\begin{sloppypar}
\noindent
S.R.~Armstrong,
E.~Charles,
P.~Elmer,
D.P.S.~Ferguson,
Y.~Gao,
S.~Gonz\'{a}lez,
T.C.~Greening,
O.J.~Hayes,
H.~Hu,
S.~Jin,
G. Mamier,
P.A.~McNamara III,
J.M.~Nachtman,$^{21}$
J.~Nielsen,
W.~Orejudos,
Y.B.~Pan,
Y.~Saadi,
I.J.~Scott,
M. Vogt,
J.~Walsh,
Sau~Lan~Wu,
X.~Wu,
G.~Zobernig
\nopagebreak
\begin{center}
\parbox{15.5cm}{\sl\samepage
Department of Physics, University of Wisconsin, Madison, WI 53706,
USA$^{11}$}
\end{center}\end{sloppypar}
}
\footnotetext[1]{Now at Harvard University, Cambridge, MA 02138, U.S.A.}
\footnotetext[2]{Permanent address: Kangnung National University, Kangnung,
Korea.}
\footnotetext[3]{Also at Dipartimento di Fisica, INFN Sezione di Catania,
Catania, Italy.}
\footnotetext[4]{Also Istituto di Fisica Generale, Universit\`{a} di
Torino, Torino, Italy.}
\footnotetext[5]{Also Istituto di Cosmo-Geofisica del C.N.R., Torino,
Italy.}
\footnotetext[6]{Now at LAL, Orsay}
\footnotetext[7]{Supported by CICYT, Spain.}
\footnotetext[8]{Supported by the National Science Foundation of China.}
\footnotetext[9]{Supported by the Danish Natural Science Research Council.}
\footnotetext[10]{Supported by the UK Particle Physics and Astronomy Research
Council.}
\footnotetext[11]{Supported by the US Department of Energy, grant
DE-FG0295-ER40896.}
\footnotetext[12]{Also at CERN, 1211 Geneva 23,Switzerland.}
\footnotetext[13]{Supported by the US Department of Energy, contract
DE-FG05-92ER40742.}
\footnotetext[14]{Supported by the US Department of Energy, contract
DE-FC05-85ER250000.}
\footnotetext[15]{Permanent address: Universitat de Barcelona, 08208 Barcelona,
Spain.}
\footnotetext[16]{Supported by the Bundesministerium f\"ur Bildung,
Wissenschaft, Forschung und Technologie, Germany.}
\footnotetext[17]{Supported by the Direction des Sciences de la
Mati\`ere, C.E.A.}
\footnotetext[18]{Supported by Fonds zur F\"orderung der wissenschaftlichen
Forschung, Austria.}
\footnotetext[19]{Supported by the US Department of Energy,
grant DE-FG03-92ER40689.}
\footnotetext[20]{Now at University of Geneva, 1211 Geneva 4, Switzerland.}
\footnotetext[21]{Now at University of California at Los Angeles (UCLA),
Los Angeles, CA 90024, U.S.A.}
\footnotetext[22]{Now at SAP AG, D-69185 Walldorf, Germany}
\footnotetext[23]{Now at D\'epartement de Physique, Facult\'e des Sciences de Tunis, 1060 Le Belv\'ed\`ere, Tunisia.}
\footnotetext[24]{Supported by the Commission of the European Communities,
contract ERBFMBICT982894.}
%centerline{(Submitted to .................)}
%
% restore the previous settings
%
\setlength{\parskip}{\saveparskip}
\setlength{\textheight}{\savetextheight}
\setlength{\topmargin}{\savetopmargin}
\setlength{\textwidth}{\savetextwidth}
\setlength{\oddsidemargin}{\saveoddsidemargin}
\setlength{\topsep}{\savetopsep}
%%%%%%%%%%%%%%%%%%%%%%%%%%%%%%%%%%%%%%%%%
\normalsize
\newpage
\pagestyle{plain}
\setcounter{page}{1}

\newpage
\pagestyle{plain}
\setcounter{page}{1}
\normalsize
\newpage
\pagestyle{plain}
\setcounter{page}{1}

\newpage
\pagestyle{plain}
\setcounter{page}{1}
%
% Main body of document 
%
\pagenumbering{arabic}
\normalsize
\setlength{\topmargin}{-1cm}
\setlength{\oddsidemargin}{-0.5cm}
\section{Introduction}
\label{sec:intro}

%Paragraph 1:
The Standard Model 
%\cite{Glashow} 
is very successful in describing the
%electroweak 
interactions of elementary particles. However, the origin of the
particle masses remains an open question. The
%putative 
Higgs mechanism
%\cite{HiggsEB} 
directly addresses this problem, with the $\W$ and $\Z$
gauge bosons acquiring mass via spontaneous symmetry breaking.
%The $\SUXU$ local gauge symmetry of the
%unified electromagnetic and weak interactions is therefore said to be
%hidden ($\mW>>0,\mathrm{m}_{\gamma}$=0).  
A consequence
of the Higgs mechanism is the addition of a neutral scalar
particle, the Higgs boson, to the spectrum of elementary particles.
%The existence of non-zero masses for fundamental fermions is explained
%through postulated $\H\ff$ couplings.
%Paragraph 2:
While the Higgs boson mass $\mH$ is not predicted by the theory,
recent results of fits to experimental electroweak data favour low
mass values
%, with $\chi^{2}(\mH)$ minima well below 200$\Gcs$
\cite{EWfits}.

%Paragraph 3:
At LEP2, the Higgs boson can be produced through the Higgs-strahlung
process, $\ee\r\H\Z$, with smaller contributions to the $\H\nn$ and
$\H\ee$ channels from W- and Z-fusion processes.  In the Standard
Model, the Higgs boson production rate and decay branching ratios are
calculable as a function of $\mH$ \cite{HiggsEB}. In the mass region
relevant to LEP2 searches, the Higgs boson decays mostly into $\bb$
and, to a lesser extent, into $\tt$.  The searches described in this
paper cover most of the topologies arising from the HZ process, with
$\H\r$ hadrons or $\tt$, and $\Z\r\ee,\mm,\tt,\nn$, or $\qq$.

%Paragraph 4:
The LEP collider has been operating at centre-of-mass energies above
the $\WW$ production threshold since 1996.  The ALEPH Collaboration
has searched for the Higgs boson with ${\sim}20\pb$ of data
accumulated at $\sqrt{s}=161$ and 170--172 GeV: no evidence of a signal
was detected, and a lower limit of $70.7 \Gcs$ at 95$\%$ confidence
level was set on the Higgs boson mass \cite{SM172}.  The other three LEP
experiments have reported similar results \cite{DELPHI172,L3172,OPAL172}.

%Paragraph 5:
A total integrated luminosity of $57\pb$ was accumulated by ALEPH in
1997 at centre-of-mass energies near 183 GeV: 
  
0.2$\pb$ at 180.8 GeV,
3.9$\pb$ at 181.7 GeV, 51.0$\pb$ at 182.7 GeV, and 1.9$\pb$ at
183.8 GeV.  
In this letter, these data are referred to as ``the 183
GeV data.''  The higher centre-of-mass energies and the larger total
luminosity increase the experimental detection sensitivity for the
Higgs boson to a mass around $85 \Gcs$.  Therefore a reference signal
with $\mH = 85 \Gcs$ and $\sqrt{s}=183 ~\G$ is used when optimizing
the event selections and quoting signal detection efficiencies.  The
total production cross section for the reference signal is 0.37 pb.

%Paragraph 6:
To improve the discriminating power between signal and background
processes, the previously published selections \cite{SM172} are
updated, and new event selections based on artificial neural networks
are introduced.  Another new feature of Higgs boson searches at
$\sqrt{s}{\sim}183~\G$ is the higher and partly irreducible $\Z\Z$
background; since these centre-of-mass energies lie on the threshold
for ZZ production, a significant fraction of these events have only
one on-shell Z boson.

%Paragraph 7:
This letter is organized in the following manner.  Section 2 provides
a brief description of the ALEPH detector and the b-tagging scheme
used in the searches.  An overview of the search strategy and of the
method used to combine the event selections in all the channels is
presented in Section \ref{sec:strategy}. The event selection criteria
for each of the signal final states
% --- $\H\ee$ and $\H\mm$, $\H\nn$, the four jet final state, 
% and final states with $\tau$ leptons --- 
are described in Sections \ref{sec:sel_hll} through
\ref{sec:sel_ttqq}; new developments with respect to the previous
publications \cite{SM172,MSSM172} are emphasized and systematic
uncertainties specific to each channel are also summarized.  In
Section \ref{sec:result} the combination of all search channels is
presented and the final result is derived.  A summary follows in
Section \ref{sec:summary}.

                 Minor edits
\section{The ALEPH detector}
\label{sec:aleph}
  
In this section the ALEPH detector parts which are 
most relevant for the
analyses that follow are succinctly described.  A more comprehensive
description of the detector and its performance is given in
Refs.~\cite{det} and \cite{perf}.

Three coaxial tracking devices surround the beam line.  The innermost
device is a silicon microstrip vertex detector (VDET)~\cite{newvdet}.
It consists of two cylindrical layers of silicon wafers situated at
average radii of 6.3 and 11.0 cm.  
%The wafers are readout in $r\phi$
%with a pitch of 50 $\mu$m and in $rz$ a pitch of 100 $\mu$m.  
Charged particles with polar angle in the range $\left|\cos
  \theta\right| < $ 0.88 traverse both VDET layers.  The VDET is
surrounded by a 2 m long inner tracking wire chamber (ITC) which
provides up to eight hits between radii of 16 and 26 cm.  Outside of
the ITC is the main tracking detector, a large time projection chamber
(TPC) which measures up to 21 three-dimensional coordinates per
charged particle.  The TPC also measures up to 338 samples of the
specific energy loss per track, allowing charged particle
identification.  A superconducting solenoid immerses the central
tracking volume in a 1.5 T axial magnetic field. 

In this letter, charged particle tracks reconstructed with at least
four hits in the TPC and which originate from within a cylinder of 2
cm radius and 20 cm length centred on the nominal interaction point
are called {\sl good tracks}.  The tracking ensemble achieves a
momentum resolution $\sigma(\pt)/\pt$ of $6 \times
10^{-4}~(\mathrm{GeV}/c)^{-1}~\pt \oplus 0.005$.  The
three-dimensional resolution on the impact parameter of tracks can be
parameterized as $(34 + 70/p)\times(1+1.6\cos^{4}\theta)~\mu$m, with
$p$~in$\Gc$.

The precise measurement of track parameters plays an essential r\^ole
in the identification of jets containing b hadrons. Weakly decaying
b hadrons are long-lived, typically flying several millimetres before
decaying.  The tagging of b-quark jets relies on six variables
reflecting the impact parameter of tracks in the jet, reconstructed
secondary vertices, identified electrons and muons with large
transverse momentum with respect to the jet axis, and jet shape and
multiplicity variables.  A neural network is used to combine this
information into an output $\eta$ for each jet.  The network is
trained to have a response near zero for light quark jets and near one for
b quark jets \cite{MSSM172}.

% Calorimetry 

A lead/proportional-chamber electromagnetic calorimeter (ECAL) is also
situated inside the superconducting coil. It is finely 
segmented into projective towers
of $0.9^{\circ} \times 0.9^{\circ}$, allowing the
identification of electrons and photons within jets.
%as well as the reconstruction of $\pi^{0}$s.  
Luminosity calorimeters are installed between the ECAL endcaps and the
beam pipe.  These calorimeters are of similar construction to ECAL and
are treated as an extension of it.
%, thus increasing the coverage of the
%electromagnetic calorimetry down to 45 mrads.  
A relative energy resolution of $0.18/\sqrt{E(\mathrm{GeV})}+0.009$ is
achieved. A silicon-tungsten sampling calorimeter completes the
electromagnetic calorimeter coverage down to 34 mrad.

The ECAL is surrounded by a hadron calorimeter (HCAL) consisting of 5
cm thick iron slabs instrumented with streamer tubes; this structure
serves as the return yoke for the magnetic field. The HCAL has the
dual purpose of measuring hadronic energy deposition as well as acting
as a filter for the identification of muons.  When used in conjunction
with the ECAL, a relative energy resolution of $0.85/\sqrt{E
(\mathrm{GeV})}$ is achieved for hadrons. The outermost detectors are two 
double layers 
of muon chambers. Muons are identified as charged
particles which penetrate the whole depth of the HCAL or which have
associated hits in the outer chambers.

The measurements of charged particle tracks and of energy deposition
in the calorimeters, combined with the identification of photons,
electrons, and muons, are used to produce a list of charged and
neutral {\sl energy flow particles} which are used in all the
analyses which follow.  Hadronic jets are clustered from these objects
with a resolution approximately described as $\sigma(E) =
(0.60\sqrt{E} + 0.6)~\mathrm{GeV} \times (1+\cos^{2}\theta)$, where
$E$ is the jet energy in GeV and $\theta$ is its polar angle.  The
resolution on the jet angles is approximately 20 mrad in both
$\theta$ and $\phi$.  

The total uncertainty on the integrated luminosity of the accumulated
data is less than 0.6\%.

\section{Search strategy}
\label{sec:strategy}

Event selections have been previously developed \cite{SM172} for the
various topologies arising from the $\H\Z$ process. These selections
address the $\H\ll$ channel (here and in the following, $\ell$ denotes
an electron or a muon,
%indistinctly called 
collectively referred to as 
``leptons''), the $\H\nn$ channel, the
$\H\qq$ channel where $\H\r\tt$ decays are not included, the $\H\tt$
channel, and the $\tt\qq$ channel which complements the $\H\qq$
channel when $\H$ decays to a $\tt$ pair. These selections are
reoptimized for the increased centre-of-mass energy and integrated
luminosity and are supplemented with new event selections.  In the
$\H\ll$ final state the selection is improved to extend the
acceptance into an angular region not considered previously.  In
final states with $\tau$ leptons, the previous track-based $\tau$
candidate selection is replaced by a minijet-based selection, and 
the previous cut-based event selection algorithms are replaced with 
new selections based upon neural network (NN) combinations of
discriminating variables.
%In final
%states with $\tau$ leptons, the previously existing parallel event
%selection algorithms are unified. 
In the $\H\nn$ and the $\H\qq$ channels, new NN-based event selections
complement the published cut-based selections \cite{SM172}.

%Paragraph 2:
The various selections are optimized to maximize the sensitivity to a
Higgs boson signal with mass $85\Gcs$, which is near the expected experimental
sensitivity.  The expected combined confidence level on the signal
hypothesis that would be obtained on average if no signal were
present~\cite{combin} is minimized with respect to the position of the
cuts on the most relevant selection variables:

\begin{itemize}
\item the reconstructed Z mass in the $\H\ll$ selection;
\item the b-quark content of the event and the event acoplanarity
in the $\H\nn$ cut-based selection;
\item the b-quark content of the jets from a candidate Higgs boson 
and the invariant mass of jets from the Z boson in the $\H\qq$
cut-based selection;
\item the neural network outputs in the $\H\nn$ and the $\H\qq$ 
neural network selections, and in the $\H\tt$ and $\tt\qq$ selections.
\end{itemize}

The confidence level for each channel is computed without performing
any background subtraction. All channels are subsequently combined using
the elitist prescription described in Ref.~\cite{combin}.  This
combination prescription assigns different weights to the various
search channels in an optimal manner according to their 
sensitivities to the signal hypothesis.

When two selections are available for a given final state, their
results are combined instead of keeping only the selection leading to
the (often marginally) better expected confidence level.  The gain
from this combination procedure is significant since, when two
analyses are aimed at selecting the same signal and have similar
performance, they often have a large overlap in signal efficiency but
not so large an overlap in terms of background.
 
Signal events are, unlike reducible background events, signal-like in
{\sl many} variables. The selection of a background event by both
selections is therefore less probable, as it depends on the
specific choice of variables in each of the selections.  The two
four-jet final state selections illustrate this point, as is
summarized in Table~\ref{tab:and}.
\begin{table}[htbp]
\caption{{\footnotesize
Signal efficiency and the number of signal and background events
expected to be selected by each of two four-jet event
selections (cuts and neural network), by the two selections
simultaneously, and by each of the two exclusively. In this table the
selection criteria of the cut-based and neural network selections 
are optimized independently of one another, and
together with all other channels.
\label{tab:and}}}
\vspace{5mm}
\begin{center}
\begin{tabular}{|l||c|rr|} 
\hline
Selection       & Efficiency ($\%$) & Signal & Background  \\ 
\hline\hline
NN              &  25.7           &  3.53   & 1.50       \\
Cuts            &  24.1           &  3.31   & 1.48       \\
\hline
Cuts and NN     &  19.8           &  2.72   & 0.88       \\
Cuts only       &   4.3           &  0.59   & 0.60       \\
NN only         &   5.9           &  0.81   & 0.62       \\
%\hline 
Total           &  30.0           &  4.12   & 2.10       \\
\hline
\end{tabular}
\end{center}
\end{table}
%Paragraph 5:
The two original selections are therefore separated into three
statistically independent and thus easily combinable sub-selections.
The first sub-selection, corresponding to the overlap of the two
original analyses, is very pure.  The two less pure exclusive
selections contain additional information and are combined with the
first, decreasing further the overall expected confidence level.  The
combination of the three sub-selections is also done with the elitist
prescription.  As a result, the events selected by both original
analyses receive a larger weight (being more signal-like) than those
selected exclusively by one of the two (being more background-like).
%

%Paragraph 3:(moved to here)
Systematic uncertainties related to the knowledge of the residual
background shape and normalization do not affect the results presented
in this letter since no background subtraction is performed.
% rm following (unnecessary?) remark:
% this is also the case for the results described in Ref.~\cite{SM172}.  
As is also the case in Ref.~\cite{SM172}, the reconstructed Higgs
boson mass for all final states is a discriminating variable 
entering the calculation of the confidence levels.
Here, the power of the confidence levels determined
from the various searches is improved
% with respect to Ref.~\cite{SM172} 
by including more discriminating variables in the
test statistic. These new variables are
\begin{itemize}
\item the b content of the event in the $\H\nn$ channel for the 
cut-based selection; 
\item the neural network output in the $\H\nn$ channel for the 
NN selection;
\item the b content of the hadronic jets for the $\H\ll$ channel.
\end{itemize}

%Paragraph 6:
A global optimization is performed in the manner described above with
the criteria of the two $\H\qq$, the two $\H\nn$, the $\H\ll$, and the
$\tt\qq$ and $\H\tt$ selections varied simultaneously.
% New 05-July-98.
The final sets of selection cuts, leading to the overall smallest
expected combined confidence level for $\mH=85\Gcs$, are described
in the following section.  
%The result of the analysis combination
%when applied to the data is presented in Section~\ref{sec:result}.

\section{Event selections}
\label{sec:sec40}

In addition to the cut-based analyses described below, NN-based 
analyses are introduced for the $\H\qq$, $\H\nn$, $\H\tt$ and
$\tt\qq$ channels to enhance the event selection performance.

Variables which alone show marginal separation between signal and
background and would be difficult to incorporate into a cut-based
analysis can be effectively used in the neural network framework~\cite{ref:NNTheory}.

In the following subsections, the distributions of simulated
background processes are normalized to the collected integrated
luminosity, and the distributions of the simulated signal are for a Higgs
boson mass of $85\Gcs$.  The background
processes were simulated as described in Ref.~\cite{SM172}, with
sample sizes typically exceeding the collected data luminosity by a
factor ${\sim}100$.  Samples of $10\,000$ signal events are simulated at
several Higgs boson mass values using the {\ttfamily HZHA} program
\cite{HZHA}.  In the case of the $\H\nn$ final state, the W-fusion
process as well as its interference with the Higgs-strahlung process
are taken into account.  Independent training and performance
evaluation samples are used for the neural network selections.

The selection efficiencies quoted in this letter are always calculated
as the fraction of events in the given channel that pass the
selection cuts.  Table
\ref{tab:eff} shows, for each event selection, the variation of efficiency 
for several Higgs boson masses as well as the
expected number of signal events for $\mH=85 \Gcs$, the expected
number of background events and the number of events selected in the
data.

\begin{table}
\caption{{\footnotesize 
Signal detection efficiencies (in percent) for different Higgs boson mass
values.  For the $\H\qq$ and $\H\nn$ channels three
independent sub-selections are used: events selected by both the
cut-based and the neural network-based selections, and events selected
exclusively by either of them.  The number of expected background
events ($n_{\mathrm{b}}$), the number of events selected in the data
($n_{\mathrm{obs}}$), and the number of expected signal events for
$\mH=85~\Gcs$ ($n_{\mathrm{s}}$) are also given.}}
\vspace{5mm}
\begin{center}
\begin{tabular}{|c||c|c|c|c|c|c|c|c|c|}
\hline
$\mH$    & $\H\ll$ & \multicolumn{3}{|c|}{$\H\nn$} & \multicolumn{3}{|c|}{$\H\qq$} & $\H\tt$ & $\tt\qq$ \\
$(\Gcs)$ &        & Cuts+NN & Cuts & NN & Cuts+NN & Cuts & NN  &       &     \\
\hline\hline
60  & 79.6 & 8.8  & 8.7 & 4.7  & 9.3  & 5.4 & 8.1 & 18.1 & 0.93 \\
70  & 77.8 & 15.8 & 4.4 & 7.9  & 12.4 & 8.6 & 9.0 & 22.9 & 2.9  \\
80  & 78.3 & 20.9 & 2.8 & 10.9 & 23.5 & 7.4 & 5.2 & 22.7 & 9.8  \\
85  & 76.2 & 17.9 & 4.0 & 12.2 & 26.1 & 7.8 & 5.2 & 20.7 & 12.6 \\
90  & 71.9 & 11.1 & 5.2 & 7.0  & 24.6 & 9.4 & 4.6 & 17.4 & 13.6 \\
95  & 31.1 & 9.5  & 2.6 & 7.2  & 19.1 & 7.6 & 6.0 & 12.7 & 11.6 \\
100 & 4.0  & 6.0  & 1.7 & 5.7  & 14.2 & 6.6 & 6.9 & 6.8  & 5.7  \\
\hline
\hline
\multicolumn{10}{|l|}{Numbers of events}\\
\hline\hline
$n_{\mathrm{b}}$ & 2.0 & 0.16 & 0.08 & 0.18 & 1.4 & 2.1      & 1.00 & 0.17 & 0.16 \\
$n_{\mathrm{obs}}$ & 3 & 0  & 0 & 0 & 1 & 2 & 1 & 0 & 0 \\
%\hline
%\multicolumn{10}{|c|}{For $\mH=85~\Gcs$} \\
%\hline
$n_{\mathrm{s}}$ & 1.1 & 0.82 & 0.18 & 0.56 & 3.6 & 1.1      & 0.71 & 0.14 & 0.16 \\
\hline
%Weight           & 2.5 & 1.0  & 2.0  & 1.2  & 1.0 & $\sim 0$ & 0.3  & 9.7  & 8.0  \\
%\hline
\end{tabular}
\label{tab:eff}
\end{center}
\end{table}

% 21.7.98: Modified. -SA
%
Although the selections differ between the various analyses, they
share common systematic uncertainties.  For the $\H\qq$, $\H\nn$, and
$\H\tt$ analyses these common systematic effects relate to the
tagging of b-quark jets.  
%Since b tagging is the most powerful
%selection criterion in both the cut-based and neural network analyses,
Systematic effects related to the simulation of track reconstruction
and to the modelling of b-hadron physics are 
the dominant source of uncertainty in both the 
cut-based and NN-based analyses.  The systematic
effect arising from the uncertainty on the b-quark fragmentation
hardness (modelled with the Peterson function \cite{ref:Peterson}) is
determined by reweighting signal events to cover the experimental
uncertainty on the parameter $\epsilon_{\mathrm b} = 0.0045\pm 0.0014$
\cite{ALEPH-epsb}.  

Differences in the b-hadron lifetimes and decay multiplicities between
the simulation and the world averages are also 
%conservatively
incorporated in the systematic uncertainty. The uncertainty due to the
modelling of the detector tracking is estimated by varying track
parameters in accordance with the experimentally measured resolution.

%Paragraph 5:
Systematic effects from possible uncertainties in the simulation of
the non-b-tagging selection variables are estimated by recomputing the
detection efficiency with reweighted signal event samples.  The
weights are determined from a direct comparison of data and simulated
background event distributions at a preselection level with suitably
large statistics.  Further studies of systematic uncertainties specific to
each selection are given below in the corresponding sections.

\subsection{The leptonic final states}
\label{sec:sel_hll}

The $\H\ll$ channel represents 6.7$\%$ of the Higgs-strahlung cross section.
%where $\ell$ refers to either an electron or a muon. 
%In what follows we will
%refer indistinctly to electrons and muons as "leptons". 
The signal events are characterized by two leptons with an invariant
mass close to $\mZ$ and a recoil mass equal to the Higgs boson
mass. Although the branching ratio of this channel is small, the
experimental signature is very clear and the Higgs boson mass can be
reconstructed with high resolution.

The event selection follows closely that of Ref.~\cite{SM172}.  To be
considered as lepton candidates charged particles must either be
identified as electrons or muons, or else must be isolated from other
particles by more than $10^\circ$. All accepted combinations of
oppositely charged lepton candidates must have at least one identified
lepton.  Mixed e-$\mu$ pairs are not considered. Final state radiation
photons from the Z boson decay products are identified and added to
isolated lepton candidates to improve the Higgs boson mass resolution.

The analysis algorithm has been simplified with respect to
\cite{SM172} regarding the treatment of events with only one
identified lepton. In such cases, the requirement on the angles of the
tracks closest to the lepton candidates as well as the requirement
that both lepton candidates be isolated have been dropped.

The definition of lepton candidates has been improved with respect to
\cite{SM172} to include leptons below the main tracking acceptance of
$\left|\cos \theta\right| \leq 0.95$.  Low angle muons are identified down to
$7.5^\circ$ from the beam line by finding hit patterns in the hadronic
calorimeter consistent with a muon. The momentum of such a low angle
muon is calculated from an over-constrained fit to the event using
energy and momentum conservation. Low angle electrons are identified
down to $12.4^\circ$ from the beam line by finding large isolated
energy deposits in the electromagnetic calorimeters. To separate these
electrons from photons, an electron candidate must have at least two
consistent hits in the ITC.  The new definition of lepton candidates
improves the $\H\ll$ signal efficiency by approximately 4$\%$ for the
same purity.

The selection criteria remain unchanged with respect to those published
in Ref.~\cite{SM172} with the sole exception of the requirement on the
reconstructed $\Z$ boson mass which is reoptimized from 
$m_{\ll(\gamma)} > 80 \Gcs$ to
$m_{\ll(\gamma)} > 82.75 \Gcs$.

\subsubsection{Systematic uncertainties}
Several potential sources of systematic uncertainty are
investigated. These include the identification of electrons and muons,
the isolation criterion, and the simulation of the energy and momentum
resolution of jets and leptons~\cite{SM172}.  Some limitations are
found in the simulation of the background to low angle electrons
which, however, do not contribute any systematic effect.  The total
relative systematic uncertainty in the signal detection efficiency is
0.4$\%$.

\subsection{The missing energy final state}
\label{sec:sel_hnn}

% Section 4.2: Paragraph 1
The $\H\Z$ signal where the $\Z$ boson decays invisibly to two
neutrinos constitutes 20$\%$ of the total Higgs-strahlung cross
section.  A large missing mass and two b jets from the Higgs boson
decay characterize this final state. Two independent analyses are used
to search for a signal in this final state: a cut-based analysis
modelled on the previous selection \cite{SM172} and an analysis based
on a neural network.

Both analyses share a common hadronic event preselection.  
 
The plane perpendicular to the thrust axis
is used to divide the event into two hemispheres.  All energy flow
particles in a hemisphere are considered to belong to a jet, and both
jets are required to have nonzero energy.  To reduce background from
$\gamma\gamma$ interactions, the preselection requires either
$P_{\mathrm{t}}>5\%\sqs$ or $\mvis>30\%\sqs$, where
$P_{\mathrm{t}}$ is the total transverse momentum of the event and
\mvis\ is the visible mass.  Finally, the missing mass of the event
must be larger than 80~$\Gcs$.  

\subsubsection{Event selection with cuts}
% Section 4.2: Paragraph 3

\label{sec:hnn_cut}
In order to reject radiative returns to the Z, the missing momentum
vector is required to point into the apparatus ($\pms>20^\circ$), and
its component along the beam line is not allowed to be large
($\left|p_{\mathrm{z}}\right|<26\Gc$).  The event acoplanarity
\cite{SM172} is used to reduce further the contamination from radiative return
events ($\ahz>0.13$). The energy deposited within a $12^\circ$ cone
around the beam axis is required to be small ($E_{12}<1.2\%\sqs$) to
reject $\mathrm{We}\nu$ and $\Z\mathrm{ee}$ events with an energetic
electron deflected at a low angle into the detector.  In order to
suppress background from $\WW\r\qq^{\prime}\tau\nu$ decays, the whole
event is reclustered into $\tau$ minijets \cite{SM172}, and the most
isolated jet is required to have a low energy ($\emj<7$~$\G$).
Finally, b-tagging information from the jets (Fig.~\ref{fig:hnn}a)
is used to enhance the signal-to-background ratio
($\eta_{1}+\eta_{2}>1.1$).

\subsubsection{Event selection with a neural network}

\label{sec:hnn_nn}

The analysis described above is complemented with a neural network
selection using the six variables of the cut selection and five
additional ones. The new variables include the fraction of the energy
deposited beyond $30^\circ$ of the beam line, the total energy within
a $30^\circ$ azimuthal wedge around the missing momentum direction, the
acollinearity of the jets, and an additional combination of the
b-tagging variables, \(\log_{10}(1-\eta_1\eta_2/2) \).  Finally,
%$\theta_{\mathrm{iso}}$, 
the isolation angle of the most isolated
track (with $p > 1 ~\Gc$) provides additional discriminating power
against $\WW\r\qq^{\prime}\tau\nu$ background.

%Section 4.2: Paragraph 6
Further details of the $\H\nn$ neural network are discussed in the 
Appendix.  Figure \ref{fig:hnn}b shows the neural network output.
Events are selected if the neural network response exceeds 0.983.

\begin{figure}[htb]
\begin{tabular}{lr}
%{\epsfig{file=hnn_btag_output.eps,width=8.0cm}}
%{\epsfig{file=hnn_nnoutput.eps,width=8.0cm}}
{\psfig{file=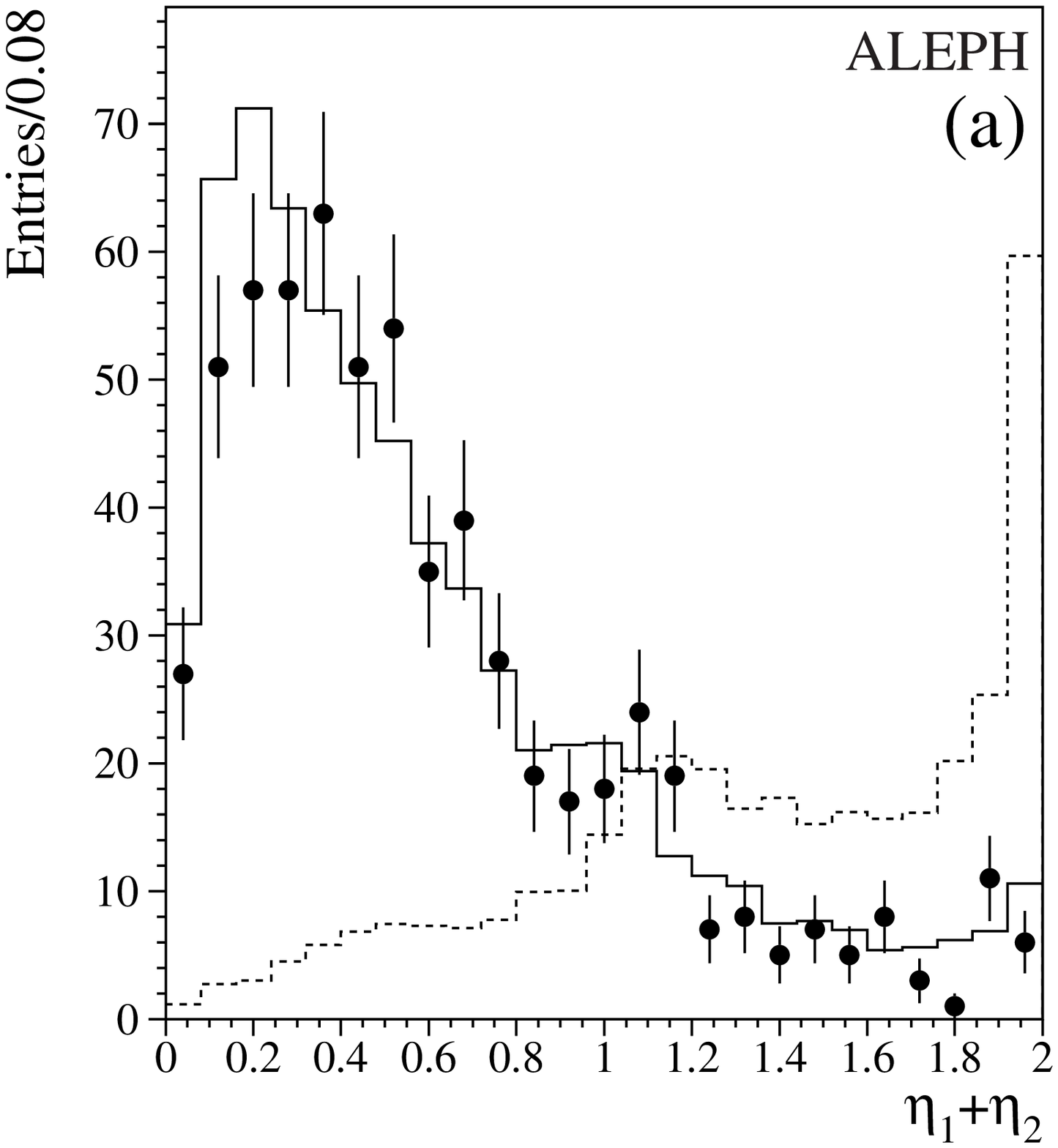,width=8.0cm}}
{\psfig{file=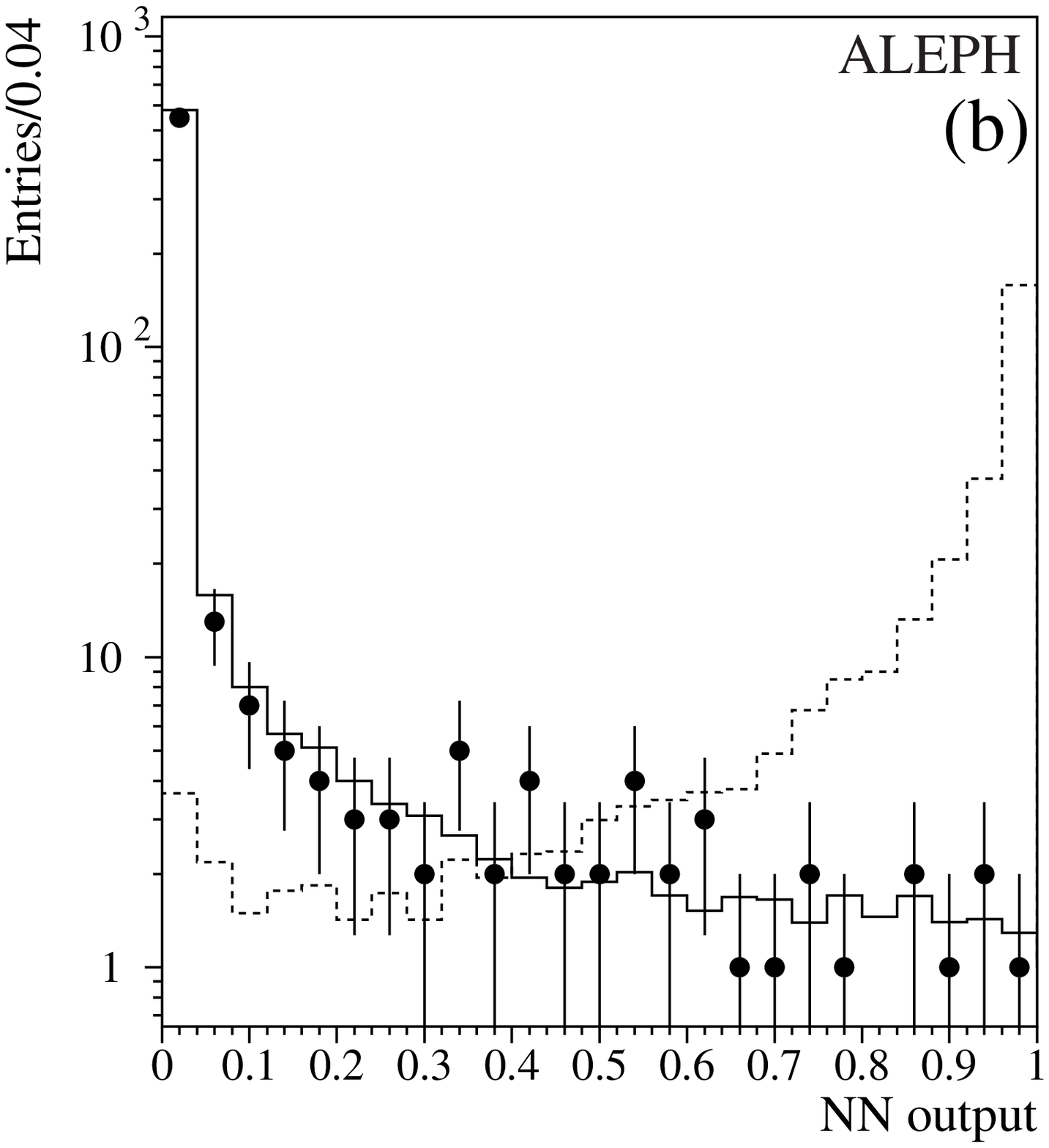,width=8.0cm}}
\end{tabular}
\caption{
{\footnotesize The distributions in the missing energy channel of (a)
the sum of the two neural network b-tag outputs and (b) the output
of the neural network used for event selection. The distributions are
shown for the data (points), the total simulated background (solid
histogram) and the reference signal sample (dashed histogram) after the
preselection cuts.  The signal distributions have an arbitrary
normalization.}
\label{fig:hnn}}
\end{figure}

\subsubsection{Systematic uncertainties}

%Section 4.2: Paragraph 10
Based on deviations between data and Monte Carlo simulation of the
kinematic variable distributions, a relative systematic uncertainty of
2.4$\%$ on the signal efficiency is determined.  The events from Monte
Carlo simulation of the HZ signal are reweighted to simulate changes
in
%the b-lifetime, b-multiplicity, and b-fragmentation, 
b fragmentation, lifetime, and decay multiplicity, 
giving a relative systematic
uncertainty of 3.0$\%$.  The relative systematic uncertainty from
discrepancies between the tracking in the simulated data and real data
is 6.1$\%$.  The total relative systematic uncertainty on the signal 
selection efficiency is 7.2\%.

\subsection{The four-jet final state}
\label{sec:sel_4j}

% Paragraph 1:
Although the $\bb\qq$ final state is not as distinctive as the
leptonic and missing energy final states, its large branching ratio,
64.4$\%$, compensates for this drawback.  The dominant background
processes $\ee\r\mathrm{q}\bar{\mathrm{q}}$, \WW, \ZZ ~are
reduced using event topology, kinematic criteria, and b-tagging
information.  Two analyses are used to search for this final state: a
cut-based analysis \cite{SM172} and a neural network based
analysis.
 
\label{sec:4j_1}
% Paragraph 2:
Common event preselection criteria are used for both the cut-based and
neural network analyses.  Events are required to have at least eight
good tracks satisfying $\left|\cos{\theta}\right| \leq 0.95$.
Radiative returns to the Z resonance are rejected when the initial state
photon is observed in the apparatus as well as when it escapes down
the beam pipe.  The events are then forced to form four jets with the
Durham jet-clustering algorithm \cite{ref:DURHAM}.  The
$y_{\mathrm{cut}}$ value where the transition from four to three jets
occurs ($y_{34}$) must be larger than~0.004.  Each jet is required to
contain at least one good track.  The energies and masses of the four
jets are then rescaled in accordance with energy and momentum
conservation.  The measured jet velocities are preserved in this
process.

\subsubsection{Event selection with cuts}
\label{sec:4j_2}

%Paragraph 3:
The analysis algorithm and the selection criteria are
identical to those found in \cite{SM172} except where explicitly
stated otherwise.

\begin{figure}[hbt] 
\begin{tabular}{lr}
%{\epsfig{file=jas.ps,width=8.0cm}} 
%& 
%{\epsfig{file=4jnn.ps,width=8.0cm}}                
{\psfig{file=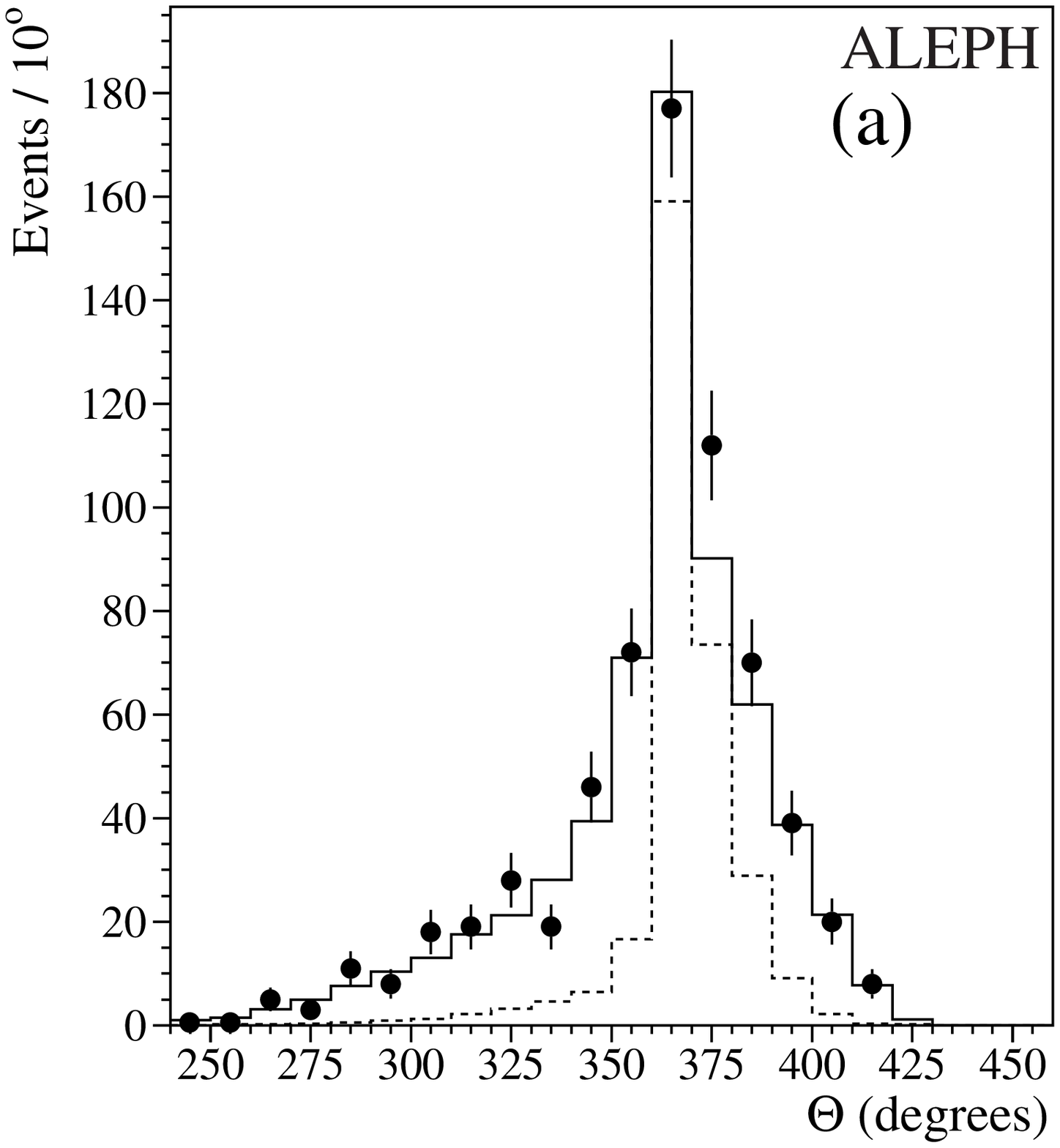,width=8.0cm}} 
& 
{\psfig{file=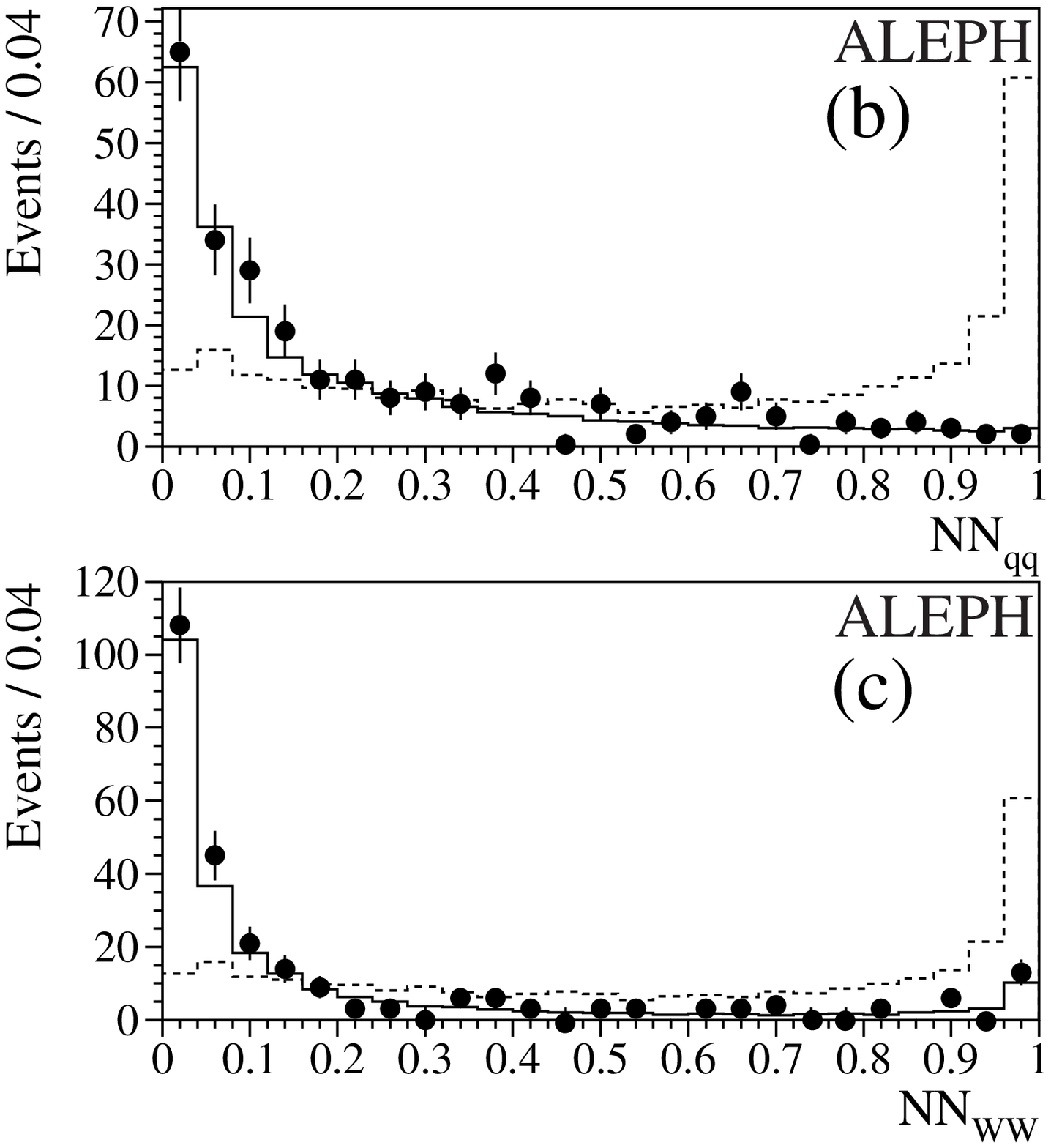,width=8.0cm}}                
\end{tabular}
\caption
{\footnotesize {The four-jet channel distributions 
for preselected events of (a)
$\Theta$, and of the outputs of (b) the anti-$\qq$ and (c) anti-$\WW$
neural networks.  The points are the data and the solid histogram is
the simulated total background normalized to the data integrated luminosity. The
signal distributions (dashed histograms) are shown with an arbitrary
normalization.}}
\label{fig:4jets} 
\end{figure}

%Paragraph 4:
The sum $\Theta$ of the four smallest interjet angles in the event (Fig.
\ref{fig:4jets}a) must be larger than $350^{\circ}$; this
significantly reduces the background contribution from
$\mathrm{q}\bar{\mathrm{q}}$ events where an energetic jet
recoils against three softer jets. Since the sensitivity of this
analysis approaches the $\H\Z$ kinematic limit, signal events are
expected to have two nearly back-to-back pairs of jets.  This event
topology is selected by requiring $\gamma =
\mathrm{min}(\cos\theta_{ij}+\cos\theta_{kl}) < -1.30$, where the minimum is
 over all possible $ijkl$ jet permutations.

%

%Paragraph 5:
Further event selection criteria are based upon the six possible
jet-pairing combinations.  An event is selected if at least one of the
jet-pairing combinations conforms to either one of the two following
sets of criteria (labelled a and b).

%Paragraph 6:
For the first case, events with four well-isolated jets are required.
One di-jet system is required to have an invariant mass consistent
with the $\Z$ mass. The other di-jet is required to have an invariant
mass within the Higgs boson mass range of interest
%\footnote{In order
%to apply this search to the Minimal Supersymmetric extension of the
%Standard Model, efficiency at low Higgs masses must be retained.}
and also to have well b-tagged jets. 
%The output of the dedicated b-tagging
%neural network for jet $i$ is denoted as $\eta_i$.  
Hence the selection criteria are the following:

\begin{description}
\item[a)] \mbox{~~~$\bullet~ y_{34} > 0.008$}; \\
$\bullet~ m_{12} > 78 \Gcs$ (Z candidate jets); \\ 
$\bullet~ m_{34} > 55 \Gcs$ (Higgs candidate jets); \\
$\bullet~ \min(\eta_3,\eta_4) > 0.30$ (Higgs candidate jets); \\
$\bullet~ (1-\eta_3)(1-\eta_4) < 6 \times 10^{-3}$ (Higgs candidate jets).
\end{description}
%Paragraph 7:
Selection criteria for the second case are designed for the
$\H\Z\r\bb\bb$ final state.  These events are almost background free and
are selected by requiring that the event as a whole has a high b-quark
content and a clear four-jet structure:
\begin{description}
\item[b)] \mbox{~~~$9.5y_{34} + \sum_{i=1}^{4} \eta_i > 2.90$.}
\end{description}

\subsubsection{Event selection with neural networks}
\label{sec:4j_3}
%Paragraph 9:
% 22.7.98: replace with PJ suggestion. -SA
Two neural networks are trained to identify the \HZS ~signal while
rejecting the $\mathrm{q}\bar{\mathrm{q}}$ and $\WW$ backgrounds:
one network is specifically designed to reject
$\mathrm{q}\bar{\mathrm{q}}$ events, while the other is designed
to reject $\WW$ events.  Details related to the training of the neural
networks are discussed in the Appendix.

% Move the following general remark to Section 4.0:
%It is possible to design and
%additional neural network to reject \ZZ events; however, studies indicate
%that the use of such a network can severely deteriorate the overall
%analysis performance for Higgs masses approaching 90 \Gcs.

%Paragraph 10:
The input patterns to the neural networks consist of the different
di-jet pairing combinations within each event.  Only pairings that
survive the preselection criteria described in Section
\ref{sec:4j_1} as well as the requirements $\sum_{i=1}^{4} \eta_i >1.0$ 
and $m_{34}>45\Gcs$ are input to the neural networks.  

The inputs for both neural networks include several of the
selection variables used in the cut-based selection: $y_{34}$, $\gamma$,
$m_{12}$, $\mathrm{min}(\eta_3,\eta_4)$, $(1-\eta_3)(1-\eta_4)$, and
$\sum_{i=1}^{4} \eta_i$.  The invariant mass of the Higgs candidate
jets $m_{34}$ is not included to avoid biasing the selection
efficiency towards the signal mass hypotheses used for training.

The average QCD four-jet matrix element squared \cite{QCDref} and
$\Theta$ 
%(described in Section \ref{sec:4j_2})
are especially powerful discriminating variables for eliminating
$\mathrm{q}\bar{\mathrm{q}}$ background events and hence are also used
as inputs to the anti-$\mathrm{q}\bar{\mathrm{q}}$ neural network.  A
vast majority of four-jet events from
$\ee\r\mathrm{q}\bar{\mathrm{q}}$ are $\qq\mathrm{gg}$.  The following
observables offer additional discriminating power between light quark
and gluon jets and are included as inputs to the
anti-$\mathrm{q}\bar{\mathrm{q}}$ neural network:
%the boosted
%aplanarity, the boosted sphericity, and the multiplicity of tracks
%with large rapidity with respect to the jet axis in the two $\Z$
%candidate jets.  
the boosted aplanarity and the boosted sphericity (both calculated 
in the rest frame of the jet), and the multiplicity of tracks
with large rapidity with respect to the jet axis, for the two $\Z$
candidate jets.  
Additional kinematic variables ({\sl e.g.}, jet
energies and di-jet masses, smallest interjet angle and angle of the
missing momentum with respect to the beam axis) are also included to
improve the overall discriminating power.  The complete list of input
variables for the anti-$\mathrm{q}\bar{\mathrm{q}}$ neural
network is given in the Appendix.

% Finally describe here the inputs specific to the anti-WW NN.
Important characteristics of $\WW$ events compared to
\HZS ~signal events include the lack of b jet
production and the comparative abundance of c jets.  
%22.7.98: Remove -SA
%The Standard
%Model Higgs boson very rarely decays to c-jets.  
In order to exploit these differences, the common neural network input
variables listed above are complemented with the b-tagging information
from the two $\Z$ candidate jets ($\eta_1$ and $\eta_2$) and with
charm rejection variables for the Higgs candidate jets ($\mu_3$ and
$\mu_4$). The charm rejection variables are computed in a similar way
to the track impact-parameter-based uds-jet probability
$P_{\mathrm{jet}}$ \cite{MSSM172} but using only 
tracks with low rapidity with respect to the jet axis
%the lowest rapidity
%tracks 
($y < 4.9$) \cite{MINUSref}.  
This rapidity criterion is effective since charmed
meson decays result in particles with transverse momentum lower than
those from b-hadron decays.  The lowest di-jet mass, the lowest jet
energy in the event, and the event broadening variable (defined in the
Appendix) are also used.  The complete list of input variables for the
anti-$\WW$ neural network is given in the Appendix.

The neural network output distributions are shown in
Figs. \ref{fig:4jets}b and \ref{fig:4jets}c for signal,
background, and data.  An event is selected when at least one of its
jet pairing combinations has neural network outputs which satisfy both
$\mathrm{NN}_{\mathrm{q}\bar{\mathrm{q}}} > 0.940$ and
$\mathrm{NN}_{\WW} > 0.964$.

\subsubsection{Systematic uncertainties}

The systematic uncertainties affecting the signal detection efficiency
are determined using the method outlined in Section
\ref{sec:sec40}. The relative uncertainty from b tagging is 6.5$\%$.
 
The systematic uncertainty corresponding to possible limitations in
the description of the most important non-b-tagging selection
variables ($y_{34},\gamma, \Theta, m_{12}, m_{34},
\mathrm{NN}_{\mathrm{q}\bar{\mathrm{q}}}$, and
$\mathrm{NN}_{\WW}$) is determined to be 2.5$\%$. 

Conservatively increasing the jet angular resolution in $\theta$ and
$\phi$ by 10$\%$ in the Monte Carlo simulation results in an
additional uncertainty of 0.8$\%$.  The correlations between all pairs
of variables used in each of the neural networks are compared between
the data and the simulated background event sample used for training;
no significant discrepancies are observed.

Adding the above contributions in quadrature, the total relative
systematic uncertainty assigned to the signal detection efficiency is
7.0$\%$.

\subsection{The final states with $\tau$ leptons}
\label{sec:sel_ttqq}

%Paragraph 1:
Two signal channels contribute to final states with at least one $\tt$
pair.  The process $\H\Z\r\H\tt$ corresponds to 3.4$\%$ of the total
Higgs-strahlung process, and $\H\r\tt, \Z\r\qq$ corresponds to an
additional 5.5$\%$.  These events are expected to have two oppositely
charged low multiplicity jets in association with missing energy from
the $\tau$ decays. The main background processes are $\Z\Z$ and $\WW$.

% 23.7.98: remove
%\subsubsection{Preselection}
\label{sec:tau_sel}
%Paragraph 2:
Multihadronic events are selected by requiring at least eight good
tracks. The total charged track energy in the event is required to exceed
20$\%$ of the centre-of-mass energy.

%Paragraph 3:
Events from radiative returns to the Z peak, \qqg, are rejected with the 
%requiring \( |P_z| + E_{\mathrm {miss}} <1.8 \gamma_{\mathrm
%{peak}} \). Here, $P_z$ and $E_{\mathrm{miss}}$ are the total
requirement $|P_z| + E_{\mathrm {miss}} <1.8 \gamma_{\mathrm
{peak}}$. Here, $P_z$ and $E_{\mathrm{miss}}$ are the total
longitudinal momentum of the event and its missing energy.  The mean
of the initial state radiation (ISR) photon spectrum peak is denoted
as $\gamma_{\mathrm {peak}}$ and is defined as $\gamma_{\mathrm
{peak}} = {\sqrt{s}}/{2} - {\mZ^2}/{2\sqrt{s}}$.  In addition, $|P_z|$
is required not to exceed $0.6
\gamma_{\mathrm{peak}}$. In order to exploit the missing energy
expected in this final state, the total missing transverse momentum of
the event is required to be at least $2.5\%\sqrt{s}$.

%Paragraph 4:
A new method for identifying $\tau$ lepton candidates replaces the
previously published track-based approach \cite{SM172}.  The event is
clustered into a large number of jets, referred to as {\sl minijets},
with mass consistent with $\mtau$.  The $\tau$ candidates are then
selected from these minijets according to the procedure described in
detail in \cite{MSSM183}.  This minijet method offers a performance
similar to the track-based selection but is simpler.

%Paragraph 5:
Events with at least two oppositely charged $\tau$ candidates are
selected.  At least one of the candidates must be single-prong.  The
remainder of the event is forced to form two jets using the Durham
clustering algorithm.  Events which have more than one $\tau$-pair
candidate yield multiple $\tt\qq$ combinations.  A $\chi^2$ is
determined for each event combination with a kinematic consistency fit
\cite{MSSM183}.  This fit requires energy-momentum conservation while
keeping the four jet directions fixed.  In the fit, the mass of the
candidate $\tau$ jets is fixed to the nominal $\tau$ lepton mass, and
the energy resolution of the two non-$\tau$ jets is taken into account.
The invariant mass of either the $\tt$ or the $\qq$ pair is
constrained to be compatible with the Z mass, depending on the
channel.  In addition, the non-$\tau$ jet momenta are constrained to be
larger than 75$\%$ of the measured value.  The combination with the
lowest $\chi^2$ is chosen and presented to the
neural networks described below.

\subsubsection{The $\H\Z\r\H\tt$ channel}
\label{sec:Httnn}
%Paragraph 6:
To discriminate between $\H\tt$ and background events, a neural
network is used with five input variables: the total missing
transverse momentum of the event, the sum of the two $\tau$ jet
isolation angles, the sum of the fitted transverse momenta of the
$\tau$ jets with respect to their nearest hadronic jet, the kinematic
consistency $\chi^2$, and
\( (\eta_1+\eta_2)\).  The jet isolation angle definition for
the second variable is the same as for the $\H\ll$ selection 
\cite{SM172}.  The last variable reflects the b-quark content of 
the two non-$\tau$ jets.  Details of the neural network are discussed in
the Appendix.

\begin{figure}[htb]
\begin{tabular}{lr}
%{\epsfig{file=htt_netout.ps,width=8.0cm}}
%{\epsfig{file=ttqq_netout.ps,width=8.0cm}}
{\psfig{file=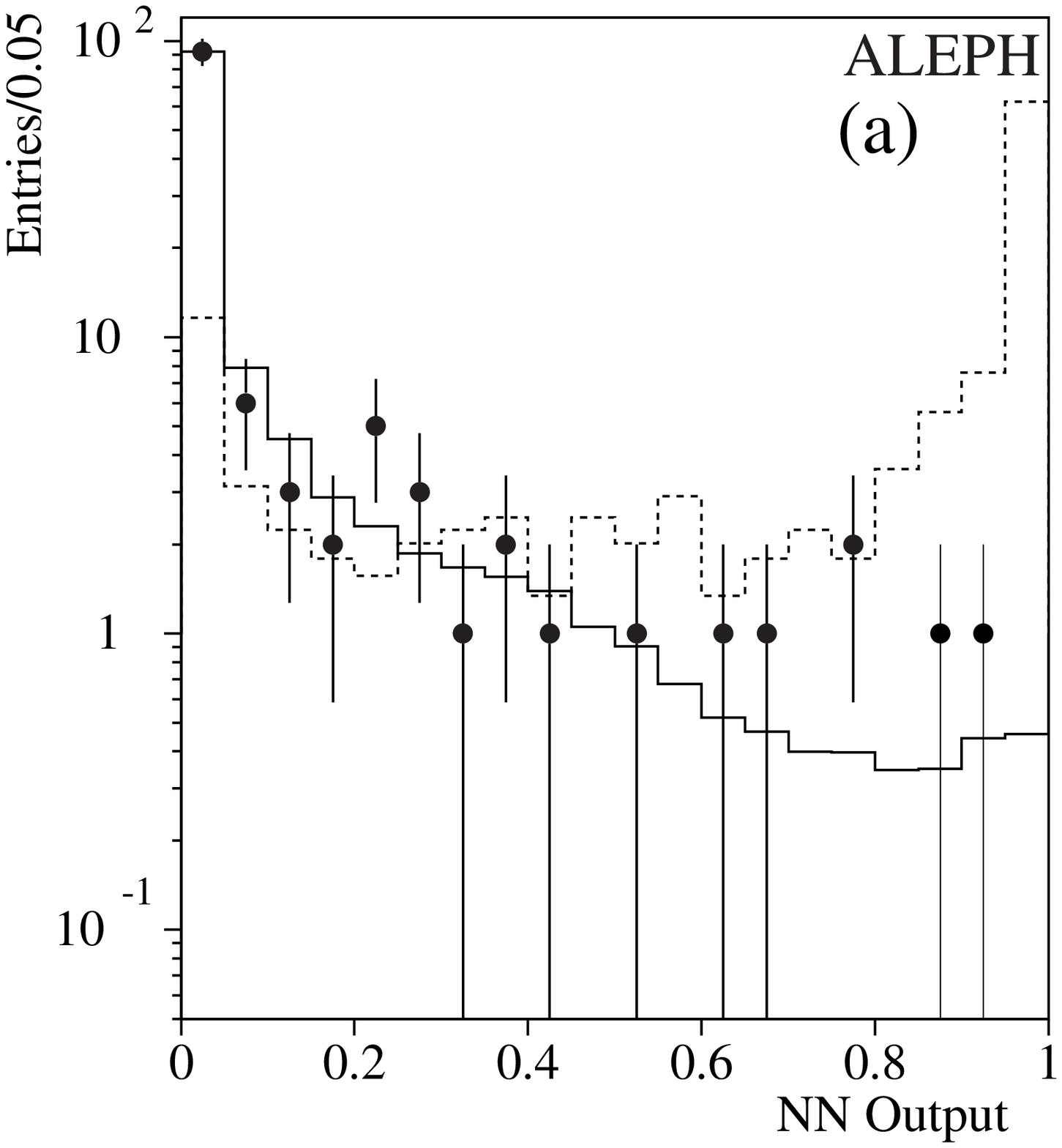,width=8.0cm}}
{\psfig{file=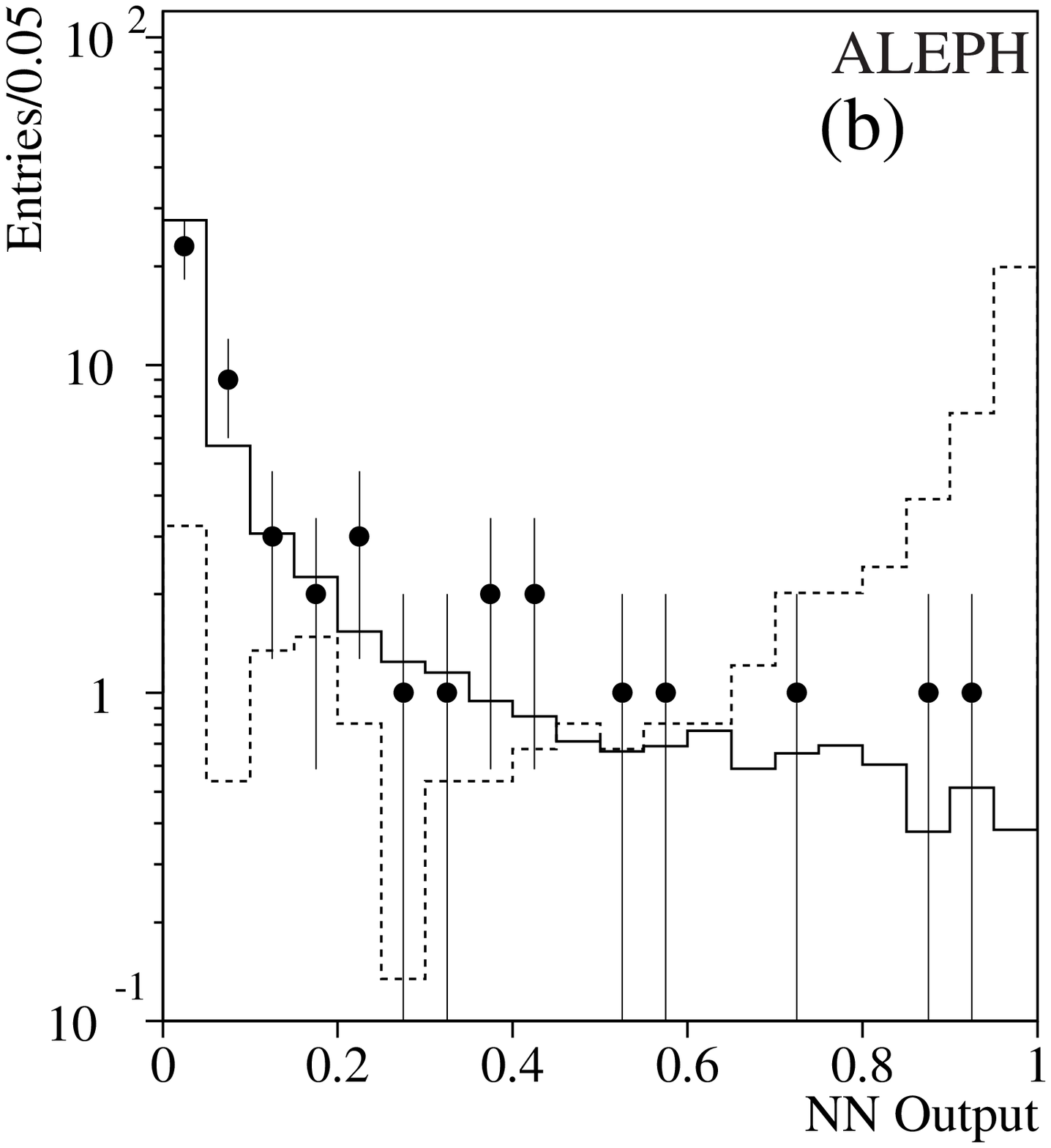,width=8.0cm}}
\end{tabular}
\caption{
{\footnotesize
The distributions of the neural network outputs for (a) the $\H\tt$
channel and (b) the $\tt\qq$ channel.  The distributions are shown for
the data (points), the total simulated background (solid histogram), and the
reference signal sample (dashed histogram) after the preselection cuts.
The signal distributions have an arbitrary normalization.}
\label{fig:ttqq}}
\end{figure}

%Paragraph 8:
Figure \ref{fig:ttqq}a shows the response of the $\H\tt$ neural
network in the data and in simulated signal and
background events.  The optimal point for this selection
corresponds to neural network output values larger than 0.979.
%yields a signal detection efficiency of 20.7$\%$.
%with 0.17 background
%events expected. No events are selected in the data.

\subsubsection{The $\H\Z\r\tt\qq$ channel}
 
%Paragraph 9:
A neural network is also used to select $\H\Z\r\tt\qq$ events.  The
input variables are identical to those
described in the previous subsection except that the b-tagging
variable is not used.  This is a limiting factor in the performance of
the selection.  Figure \ref{fig:ttqq}b shows the neural network output
in the data and simulated signal and background events.
%Therefore an additional selection criteria is added: the $\tt$
%invariant mass is required to be in excess of $50 \Gcs$.  
Events which have a neural network output larger than 0.980 are
selected.
% resulting in a signal selection efficiency of 12.6$\%$. 
%No
%events are selected in the data, in agreement with the expectation of
%0.16 events from standard processes.

\subsubsection{Systematic uncertainties}
The main sources of systematic uncertainty in the final states with
$\tau$ leptons are the reconstruction of the kinematic variables, the
clustering of jets and the tagging of jets containing b quarks.

From a comparison of multihadronic events with an identified lepton,
in the data and in simulated events, a slight discrepancy in the total
transverse momentum distribution 
is observed which translates into an absolute
uncertainty of 0.3$\%$ (0.2$\%$) in the signal detection efficiency
for the $\H\tt$ ($\tt\qq$) channel. The effects from jet
reconstruction are evaluated by smearing the
reconstructed jet angles and energies according to the expected
measurement errors. The systematic effect on the detection efficiency
is 0.4$\%$ (0.1$\%$) from the jet directions and 0.1$\%$ (0.1$\%$)
from the jet energies, in absolute terms. Uncertainties arising from
the tagging of b-quark jets affect only the $\H\tt$ selection and are
evaluated, as is described in Section~\ref{sec:sec40}, to be
0.9$\%$. Therefore, total relative systematic uncertainties of 5.0$\%$
and 1.9$\%$ are respectively assigned to the $\H\tt$ and $\tt\qq$
signal detection efficiencies.

%
% Combined Result
%---------------------------------------\input{results.tex}
\section{Combined result}
\label{sec:result}

\begin{figure}[htb]
\hspace{0.25\textwidth}
%\mbox{\epsfig{file=combrecomass.eps,width=8.0cm}}
\mbox{\psfig{file=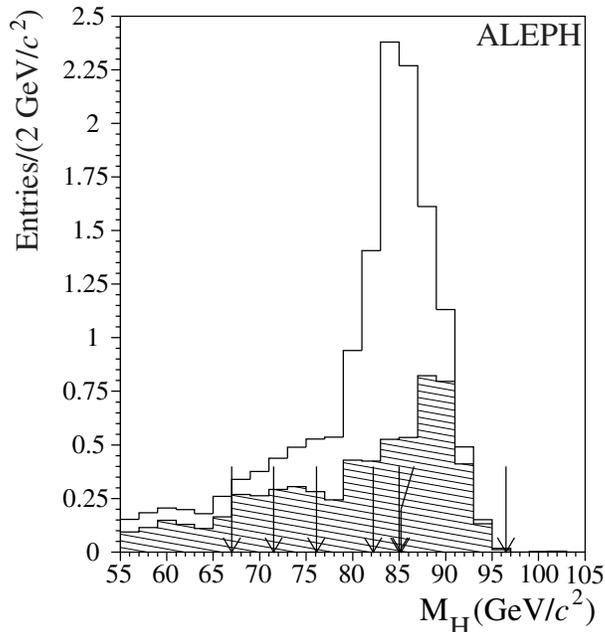,width=8.0cm}}
\vspace{-5mm}
\caption
{\footnotesize{The distribution of the reconstructed Higgs boson mass
of the selected events in all search channels. The histogram shows the
expected distribution, with the contribution from background processes
(shaded) and from the reference signal ($\mH=85\Gcs$), normalized to
the collected data luminosity. The data events are indicated with
arrows. The selected events do not have equal weights in the
combination procedure but have been plotted here with unit weight for
convenience.}}
\label{fig:mHiggs}
\end{figure}

Seven events are selected in the data by the various selections, in
agreement with the 7.2 events expected from all Standard Model
background processes.  Four of these events are selected in the
four-jet final state.  One event, selected by both the selection with
cuts and the selection with neural networks, has a reconstructed Higgs
boson mass of 71.5\Gcs\ while the other events selected with cuts
(neural networks) only, have mass values of 76.1 and 85.0\Gcs~(85.2\Gcs). 
The other three candidate events are found in the leptonic
final state, with masses of 67.0, 82.2, and 96.5\Gcs. Figure
\ref{fig:mHiggs} shows the reconstructed Higgs boson mass distribution
for the selected events in all channels.

\begin{figure}[htb]
\begin{picture}(160,85)
%\put(0,-3){\epsfxsize88mm\epsfbox{hz183.ps}}
\put(38,-3){\epsfxsize100mm\epsfbox{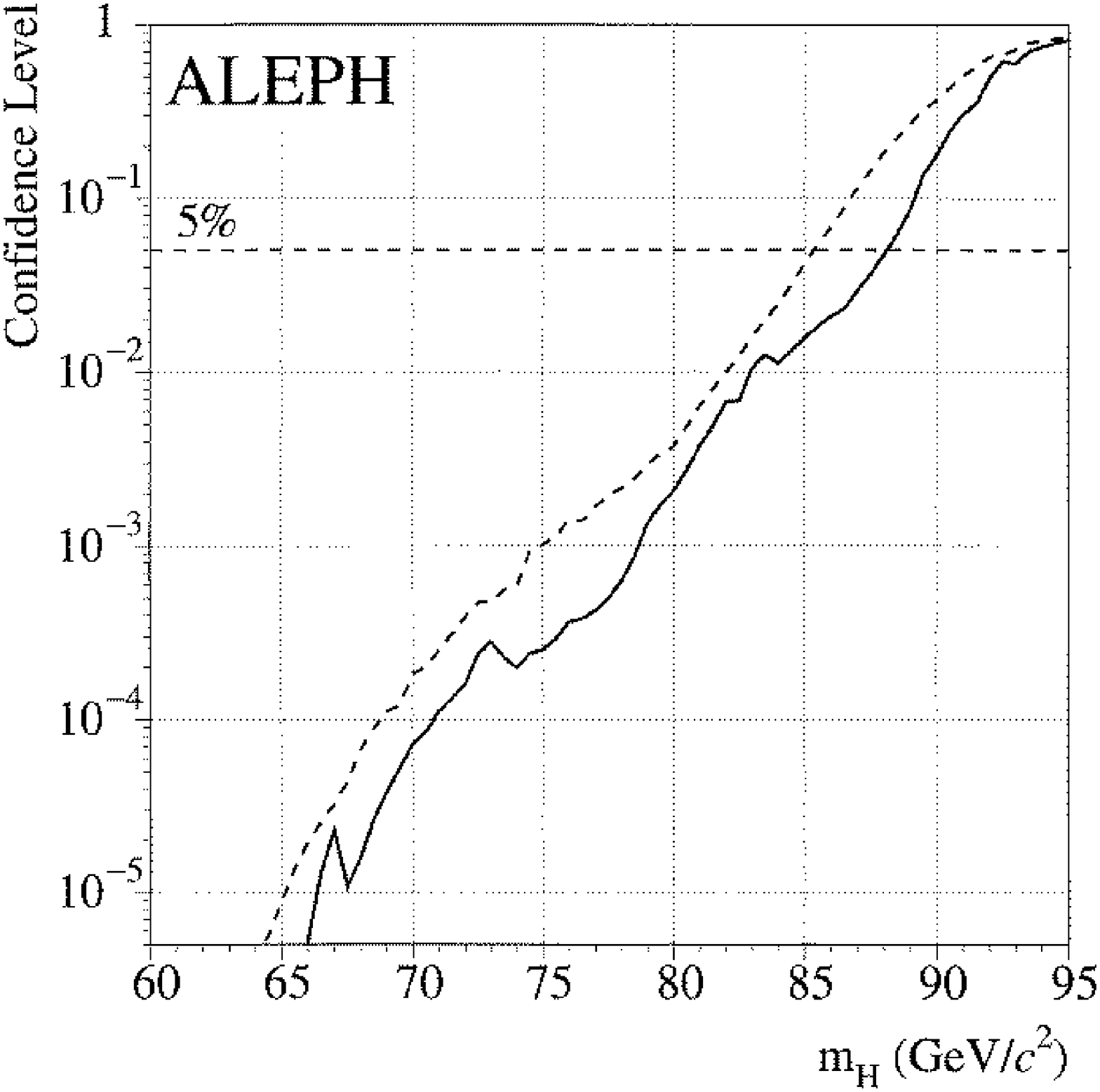}}
\end{picture}
\caption{
\footnotesize{
The observed (solid line) and expected (dashed line) confidence level
curves after all channels are combined, using the 183 GeV data as well
as the results obtained at lower energies.
\label{fig:comb183}}}
\end{figure}

%The expected and observed confidence levels in the signal hypothesis
%are in satisfactory agreement in all the channels, as shown in
%Fig.~\ref{fig:channel-limits}. 
The result obtained when all sub-selections are combined is displayed
in Fig.~\ref{fig:comb183}.
%The observed confidence level is again in
%fair agreement with the expectation over the whole mass range. 
This result includes the lower energy ALEPH data taken at
LEP1~\cite{SMLEP1} and at $\sqrt{s} = 161, 170$, and
172~GeV~\cite{SM172}.  The lower energy data has an impact on the
confidence level only for mass values lower than $\sim 75\Gcs$.

The HZ searches exclude all Higgs boson masses below 88.0\Gcs\ at the
95$\%$ confidence level. The average limit expected in the absence of
signal is 85.3\Gcs. With this expected limit, the probability to
observe a limit at least as high as 88.0\Gcs\ is 22$\%$.

As explained in the previous sections, a number of possible systematic
effects are studied and their impacts evaluated. The uncertainties
related to the selection procedure and to the inadequacies of the
Monte Carlo dominate, translating into uncertainties on the selection
efficiencies ranging from $\sim 0.5\%$ to 7$\%$, depending on the final state.
Following the method of Ref.~\cite{cousin}, a small increase of the
confidence level results, corresponding to a change in the mass limit
of $-0.1\Gcs$.

The mass of the Standard Model Higgs boson therefore exceeds 87.9\Gcs\
at the 95$\%$ confidence level. Had background subtraction been
performed --- with the same set of selection criteria and ignoring the
related systematic uncertainties --- a 95$\%$ confidence level limit of
88.3\Gcs\ would have been derived, with 88.3\Gcs\ expected.
%\cite{ref:shanmcnamara} with the same set of selection criteria.

\section{Summary}
\label{sec:summary}

A search for evidence of the production of Higgs bosons in $\ee$
collisions at centre-of-mass energies between 181 and 184 GeV has been
performed with the ALEPH detector.  The major event topologies have
been covered: the leptonic and missing energy final states as well as
the final states with four hadronic jets and those including $\tau$
leptons. The previously published search algorithms have been improved
and new selections using artificial neural networks were introduced.

% 7 events observed, limit includes the previous data collected by ALEPH;

In the collected data sample, corresponding to a total of $57\pb$,
seven events were selected, in agreement with 7.2 events expected from
Standard Model background processes.  From this observation, a 95$\%$
confidence level lower limit on the mass of the Higgs boson is set at
87.9\Gcs. A similar result has been reported by the L3 experiment at
LEP~\cite{L3_183}.

%
%---------------------------------------\input{acknowledgement.tex}
\section*{Acknowledgements}

We wish to congratulate our colleagues from the accelerator divisions 

for the very successful operation of LEP at high energy.  We are 

indebted to the engineers and technicians in all of our 

institutions for their contribution to the excellent performance

of ALEPH.  Those of us from non-member countries thank CERN

for its hospitality.
%
% Appendix
%---------------------------------------\input{appendix.tex}
\pagebreak
\begin{appendix}
\section*{Appendix}
%\label{NNApp}
This appendix offers more details on the neural networks
used for the $\H\nn$, $\H\qq$, $\H\tt$, and $\tt\qq$
selections.  All of the neural networks are fully-connected multi-layer
feed-forward networks trained with back-propagation algorithms
using existing programs \cite{ref:SNNSJETNETMODULAD}.  
During an initial training phase, patterns composed of sets of
discriminating variables are presented to the neural network, and the
network weights are optimized to produce an output near one for signal
patterns and near zero for background patterns.  

The architecture and training details of the neural networks
are summarized in Table \ref{tab:archdet}.  These parameters
are chosen based upon optimal performance in Monte Carlo
simulation.  Each network is trained with Monte Carlo
simulation of HZ signal samples as well as samples of
background processes.
\begin{table}[htb]
\begin{center}
\caption{
\footnotesize{A summary of the architecture and
the HZ signal and background process training samples used for
the neural networks described in the text.}}
\label{tab:archdet}
\vspace{5mm}
\begin{tabular}{|l|c|c|c|}\hline
Final & Network & $\mH$ in Signal & Background \\
State   & Architecture & Training Samples (\Gcs) & Training Samples\\
\hline \hline
$\H\qq$ (anti-$\mathrm{q}\bar{\mathrm{q}}$) & 22-10-10-1 & 70, 80, 85 & $\qqg$\\
$\H\qq$ (anti-\WW) & 14-12-10-2 & 80 & $\WW$\\
$\H\nn$ & 11-20-3 & 80, 85 & $\WW$, $\qqg$\\
$\H\tt$ & 5-10-1  & 80, 85, 90 & $\WW$, ZZ\\
$\tt\qq$ & 4-10-1 & 80, 85, 90 & $\WW$, ZZ\\
\hline
\end{tabular}
\end{center}
\end{table}

For the four-jet final state, only the correct pairing in \HZS~ events
is presented to the neural network as the signal pattern during the
training phase.

The input variables used for the anti-$\qq$ neural network in
the $\H\qq$ channel are listed in Table \ref{tab:nnqq}.

\begin{table}[htbp]
\begin{center}
\caption{
\footnotesize{The complete list of inputs to the anti-$\qq$ neural network. 
The variables marked with $\dagger$ are common to the NN-based and
cut-based selections and are described in Sections \ref{sec:4j_1} and
\ref{sec:4j_2}.\label{tab:nnqq}}}
\vspace{5mm}
\begin{tabular}{|rl|}\hline
Variable & \\ \hline
1. & $ y_{34}~\dagger $ \\
2.& $\gamma~\dagger $ \\
3.& $ m_{12}~\dagger $ \\
4.& min$(\eta_3,\eta_4)~\dagger $ \\
5.& $(1-\eta_3)(1-\eta_4)~\dagger $ \\
6.& $ \sum_{i=1}^{4} \eta_i~\dagger $ \\
7.& $\Theta~\dagger $ \\
8.& $\langle \mathcal{M} \rangle$, average four-jet QCD matrix element.\\
9--10. & Boosted aplanarity of $\Z$ candidate jets.\\
11--12. & Boosted sphericity of $\Z$ candidate jets.\\
13--14. & Multiplicity of tracks with rapidity
larger \\
 & than 1.6, in the $\Z$ candidate jets. \\
15. & $\min (\theta_{ij})$, lowest interjet angle. \\
16. & $\cos \pms$, cosine of the polar angle of the \\
 & event missing momentum vector. \\
17. & $\min(m_{ij}+m_{kl})$, lowest di-jet mass sum.\\
18--19. & $m_{\mathrm{min}}, m_{\mathrm{min2}}$, the two lowest jet masses.\\
20--21. & $E_{\mathrm{min}}, E_{\mathrm{min2}}$, the two lowest jet energies.\\
22. & $E_{\mathrm{max}}$, largest jet energy. \\
\hline
\end{tabular}
\end{center}
\end{table}

The input variables used for the anti-\WW neural network in the
$\H\qq$ channel are listed in Table \ref{tab:nnww}.  The event
broadening is computed using the two event hemispheres defined by the
plane perpendicular to the event thrust axis; the quantity $B_{\mathrm{hemi}}$ is computed for each hemisphere:
\[ 
 B_{\mathrm{hemi}} = \frac{\displaystyle
 \sum_{i=1}^{N_{\mathrm{tracks}}} \mid p_{\mathrm{t}i} \mid}
 {\displaystyle \sum_{i=1}^{N_{\mathrm{tracks}}} \mid p_{i}
 \mid}\, ,  \]
 where $N_{\mathrm{tracks}}$ is the number of tracks in the
 hemisphere, $p_{i}$ is the momentum of the $i$th track, and
 $p_{\mathrm{t}i}$ is the transverse momentum of the $i$th track with
 respect to the thrust axis.  The event broadening $B$ is the smaller 
 of the two $B_{\mathrm{hemi}}$ values.
\vspace{-4mm}
\begin{table}[htbp]
\begin{center}
\caption{
\footnotesize{The complete list of inputs to the anti-$\WW$ neural network. 
The variables marked with $\dagger$ are common to the NN-based and
cut-based selections and are described in Sections \ref{sec:4j_1} and
\ref{sec:4j_2}.\label{tab:nnww}}}
\vspace{5mm}
\begin{tabular}{|rl|}\hline
Variable & \\ \hline
1. & $ y_{34}~\dagger $ \\
2.& $\gamma~\dagger $ \\
3.& $ m_{12}~\dagger $ \\
4--5.& $\min(\eta_3,\eta_4), \max(\eta_3,\eta_4) ~\dagger $ \\
6.& $(1-\eta_3)(1-\eta_4)~\dagger $ \\
7.& $ \sum_{i=1}^{4} \eta_i~\dagger $ \\
8--9.& $\min(\eta_1,\eta_2), \max(\eta_1,\eta_2)$\\
10--11.& $\min(\mu_3,\mu_4), \max(\mu_3,\mu_4)$ \\
12.& $B$, event broadening. \\
13.& $E_{\min}$, lowest jet energy. \\
14.& $\min(m_{ij})$, lowest di-jet mass.\\
\hline
\end{tabular}
\end{center}
\end{table}
\vspace{-1cm}

\end{appendix}

%
% Bibliography
%---------------------------------------\input{biblio.tex}
\pagebreak

\def\PL#1#2#3{{ Phys. Lett. }{\bf B#1 }(#2) #3}
\def\NPB#1#2#3{{ Nucl. Phys. }{\bf B#1 }(#2) #3}
\def\NPA#1#2#3{{ Nucl. Phys. }{\bf A#1 }(#2) #3}
\def\PRL#1#2#3{{ Phys. Rev. Lett. }{\bf #1 }(#2) #3}
\def\PRD#1#2#3{{ Phys. Rev. }{\bf D#1 }(#2) #3}
\def\PRep#1#2#3{{ Phys. Rep. }{\bf #1 }(#2) #3}
\def\ZPC#1#2#3{{ Z. Phys. }{\bf C#1 }(#2) #3}
\def\EPJ#1#2#3{{ E. Phys. J. }{\bf C#1 }(#2) #3}
\def\NIM#1#2#3{{ Nucl. Instrum. Methods}
{\bf A#1 }(#2) #3}
\def\CPC#1#2#3{{ Comput. Phys. Commun. }
{\bf #1 }(#2) #3}
\def\etal{et al.}

\def\Aleph{ALEPH Collaboration, }
\def\AlephB{ALEPH Collaboration, }
\def\AlephC{ALEPH Collaboration, }
\def\Al#1#2#3{\Aleph \PL{#1}{#2}{#3}}
\def\AlB#1#2#3{\AlephB \PL{#1}{#2}{#3}}

\def\np#1#2#3{           {Nucl. Phys. }{\bf #1} (19#2) #3}
\def\npp#1#2#3{       {\it Nucl. Phys. (Proc. Suppl.) }{\bf #1} (19#2) #3}
\def\pl#1#2#3{           {Phys. Lett. }{\bf #1} (19#2) #3}
\def\pr#1#2#3{           {\it Phys. Rev. }{\bf #1} (19#2) #3}
\def\prep#1#2#3{         {\it Phys. Rep. }{\bf #1} (19#2) #3}
\def\prl#1#2#3{          {Phys. Rev. Lett. }{\bf #1} (19#2) #3}
\def\zp#1#2#3{           {\it Zeit. f\"ur Physik }{\bf #1} (19#2) #3}

\end{document}